\documentclass[12pt]{emulateapj}

\usepackage{natbib, graphics}

\def\umy{\hbox{\it u--y\/}}
\def\umj{\hbox{\it u--J\/}}
\def\bmj{\hbox{\it b--J\/}}

\def\vmb{\hbox{\it v--b\/}}
\def\vmk{\hbox{\it v--K\/}}

\def\vmy{\hbox{\it v--y\/}}
\def\bmy{\hbox{\it b--y\/}}
\def\bmh{\hbox{\it b--H\/}}
\def\bmv{\hbox{\it B--V\/}}
\def\jmk{\hbox{\it J--K\/}}
\def\ymk{\hbox{\it y--K\/}}
\def\feh{\hbox{\rm [Fe/H]}}

\def\Min{${}^{\prime}$\llap{.}}
\def\Sec{${}^{\prime\prime}$\llap{.}}

\def\min{${}^{\prime}$}

\def\muno{\hbox{\rm $m_1$}}

\def\afe{\hbox{\rm [$\alpha$/Fe]}}
\def\Sec{${}^{\prime\prime}$\llap{.}}
\def\min{${}^{\prime}$}
\def\Min{${}^{\prime}$\llap{.}}

\newcommand{\msun}{$M_{\odot}\,$}

\newcommand{\strom}{\mbox{Str\"omgren~}}
\newcommand{\omc}{\mbox{$\omega$ Cen~}} 
\newcommand{\omcp}{\mbox{$\omega$ Cen}}

\newcommand{\omtre}{\hbox{$\,\omega$\,3}}

\shorttitle{\strom photometry of GGCs. II Metallicity distribution of RGs in \omcp.} 
\shortauthors{Calamida et al.}

\begin{document}



\title{Str\"omgren photometry of Galactic Globular Clusters. II Metallicity
distribution of red giants in $\omega$ Centauri.\altaffilmark{1}}

\author{
A. \ Calamida\altaffilmark{2,3,4},
G. \ Bono\altaffilmark{3,4},
P. \ B. \ Stetson\altaffilmark{5},
L. \ M. \ Freyhammer\altaffilmark{6},
A. \ M. \ Piersimoni\altaffilmark{7},
R. \ Buonanno\altaffilmark{3,4},
F. \ Caputo\altaffilmark{4}, 
S. \ Cassisi\altaffilmark{7}, 
M. \ Castellani\altaffilmark{4}, 
C. \ E. \ Corsi\altaffilmark{4},
M. \ Dall'Ora \altaffilmark{10},
S. \ Degl'Innocenti\altaffilmark{11,12}, 
I. \ Ferraro\altaffilmark{4}, 
F. \ Grundahl\altaffilmark{8},
M. \ Hilker\altaffilmark{2},
G. \ Iannicola\altaffilmark{4},
M. \ Monelli\altaffilmark{13},
M. \ Nonino\altaffilmark{14},
N. \ Patat\altaffilmark{4}, 
A. \ Pietrinferni\altaffilmark{7},
P. \ G. Prada Moroni\altaffilmark{11,12},
F. \ Primas\altaffilmark{2}, 
L. \ Pulone\altaffilmark{4},
T. \ Richtler\altaffilmark{9},
M. \ Romaniello\altaffilmark{2},
J. \ Storm\altaffilmark{15},
A.\ R. \ Walker\altaffilmark{16}
}

\altaffiltext{1}{
Based on observations collected in part with the 1.54m Danish Telescope
and with the NTT@ESO Telescope operated in La Silla, and in part
with the VLT@ESO Telescope operated in Paranal. The \strom data were 
collected with DFOSC2@Danish (proprietary data), while the NIR data 
were collected with SOFI@NTT, proposals: 66.D-0557 and 68D-0545 
(proprietary data), 073.D-0313 and 59.A-9004 (ESO Science archive) 
and with ISAAC@VLT, proposal 075.D-0824 (proprietary data).
}
\altaffiltext{2}{European Southern Observatory, Karl-Schwarzschild-Str. 2,
D-85748 Garching bei Munchen, Germany; acalamid@eso.org, mhilker@eso.org, 
fpatat@eso.org, fprimas@eso.org, mromanie@eso.org 
}
\altaffiltext{3}{Universita' di Roma Tor Vergata, Via della Ricerca Scientifica 1,
00133 Rome, Italy; bono@roma2.infn.it, buonanno@roma2.infn.it
}
\altaffiltext{4}{INAF-Osservatorio Astronomico di Roma, Via Frascati 33, 
00040, Monte Porzio Catone, Italy; caputo@mporzio.astro.it, 
m.castellani@mporzio.astro.it, corsi@mporzio.astro.it, 
ferraro@mporzio.astro.it, giacinto@mporzio.astro.it, 
pulone@mporzio.astro.it 
}
\altaffiltext{5}{
Dominion Astrophysical Observatory, Herzberg Institute of Astrophysics,
National Research Council, 5071 West Saanich Road, Victoria, BC V9E~2E7,
Canada; Peter.Stetson@nrc-cnrc.gc.ca
}
\altaffiltext{6}{
Centre for Astrophysics, University of Central Lancashire,
Preston PR1 2HE;  lfreyham@googlemail.com  
}
\altaffiltext{7}{
INAF-Osservatorio Astronomico di Collurania, via M. Maggini,
64100 Teramo, Italy; cassisi@te.astro.it, piersimoni@te.astro.it, 
adriano@te.astro.it
}
\altaffiltext{8}{
Department of Physics and Astronomy, Aarhus University,
Ny Munkegade, 8000 Aarhus C, Denmark; fgj@phys.au.dk
}
\altaffiltext{9}{
Universidad de Concepcion, Departamento de Fisica, Casilla 
106-C, Concepcion, Chile; tom@coma.cfm.udec.cl
}
\altaffiltext{10}{
INAF-Osservatorio Astronomico di Capodimonte, 
via Moiariello 16, 80131 Napoli, Italy; dallora@na.astro.it  
}
\altaffiltext{11}{
Dipartimento di Fisica "E. Fermi", Univ. Pisa,
Largo B. Pontecorvo 3, 56127 Pisa, Italy; prada@df.unipi.it, scilla@df.unipi.it
}
\altaffiltext{12}{
INFN, Sez. Pisa, via Largo B. Pontecorvo 2, 56127 Pisa, Italy
}
\altaffiltext{13}{
IAC - Instituto de Astrofisica de Canarias, Calle Via Lactea,
E38200 La Laguna, Tenerife, Spain; monelli@iac.es
} 
\altaffiltext{14}{
INAF-Osservatorio Astronomico di Trieste, via G.B. Tiepolo 11,
40131 Trieste, Italy; nonino@ts.astro.it  
} 
\altaffiltext{15}{
Astrophysikalisches Institut Potsdam, An der Sternwarte 16,
D-14482 Potsdam, Germany; jstorm@aip.de
} 
\altaffiltext{16}{
Cerro Tololo Inter-American Observatory, Casilla 603, 
La Serena, Chile; awalker@noao.edu
} 

\date{\centering drafted \today\ / Received / Accepted }

\begin{abstract}
We present new intermediate-band \strom photometry based on
more than 300 $u,v,b,y$ images of the Galactic globular cluster
\omcp. Optical data were supplemented with new multiband near-infrared
(NIR) photometry (350 $J,H,K_s$ images). The final optical-NIR catalog
covers a region of more than $20\times20$ arcmin squared across the
cluster center. We use different optical-NIR color-color planes together
with proper motion data available in the literature to identify candidate
cluster red giant (RG) stars. By adopting different \strom metallicity
indices we estimate the photometric metallicity for $\approx 4,000$
RGs, the largest sample ever collected. The metallicity distributions
show multiple peaks ($\feh_{phot}=$ $-1.73\pm0.08$, $-1.29\pm0.03$,
$-1.05\pm0.02$, $-0.80\pm0.04$, $-0.42\pm0.12$ and $-0.07\pm0.08$ dex)
and a sharp cut-off in the metal-poor tail ($\feh_{phot} \lesssim -2$ dex)
that agree quite well with spectroscopic measurements.
We identify four distinct sub-populations,
namely metal-poor  (MP, $\feh \le -1.49$),  metal-intermediate
(MI, $-1.49 < \feh \le -0.93$),  metal-rich  (MR, $-0.95 < \feh \le -0.15$)
and solar metallicity (SM, $\feh \approx 0$). The last group includes only a
small fraction of stars ($\sim\; 8\pm5$\%) and should be confirmed
spectroscopically. 
Moreover, using the difference in metallicity based on
different photometric indices, we find that the $19\pm1$\% of RGs
are candidate {\rm CN}-strong stars. This fraction agrees quite well with
recent spectroscopic estimates and could imply a large fraction of
binary stars. The \strom metallicity indices display a robust
correlation with $\alpha$-elements ({\rm [Ca+Si/H]}) when moving
from the metal-intermediate to the metal-rich regime
($\feh \gtrsim -1.7$ dex).
\end{abstract}



\section{Introduction}\label{introduction}

The Galactic Globular Cluster (GGC) \omc (NGC\,5139) is 
currently the target of significant observational efforts covering 
the whole wavelength spectrum. This huge star cluster, the most 
massive known in our Galaxy (2.5 $\times 10^6$ \msun, van de Ven et al.\ 2006)
hosts at least three separate stellar populations with a large 
undisputed spread in metallicity (Norris \& Da Costa 1995; 
hereinafter ND95; Norris et al.\ 1996, hereinafter N96; 
Suntzeff \& Kraft 1996, hereinafter SK96; Smith et al.\ 2000, hereinafter SM00; 
Kayser et al.\ 2006, hereinafter KA06; Villanova et al.\ 2007).
The most relevant morphological and structural parameters of \omc are 
summarized in Table~1.

The unusual spread in color of the red giant branch (RGB) in \omc was revealed 
for the first time by Dickens \& Woolley (1967) using photographic photometry 
by Woolley (1966), 
and later verified by  Cannon \& Stobie (1973) using photoelectric photometry.
The width in color of the RGB was interpreted as an indication of an 
intrinsic spread in chemical abundance, as subsequently confirmed by 
spectroscopic data (Freemann \& Rodgers 1975). Moreover, recent photometric 
surveys by Lee et al.\ (1999), Pancino et al.\ (2000, hereinafter PA00), 
Rey et al.\ (2004) and Sollima et al.\ (2005a, hereinafter S05a), disclosed 
the discrete nature of the \omc RGBs.

More recent spectroscopic studies of red-giant (RG) stars have refined our
knowledge of the intrinsic spread in heavy elements in \omc (ND95; N96; 
SK96). In particular, the low-resolution study of 
$\approx$ 500 RGs performed by N96 and by SK96, based on the 
{\rm Ca} $H$, $K$ lines, and on the infrared {\rm Ca}{\footnotesize\rm II} triplet (CaT), 
provided a metallicity distribution with a dominant peak located 
at \feh\ $\approx$--1.6, a secondary peak at \feh\ $\approx$--1.2, 
and a long, asymmetric tail extending toward higher metallicities 
(\feh\ $\approx$--0.5). This distribution agrees quite well with 
the metallicity distributions obtained by Hilker \& Richtler (2000, 
hereinafter HR00, $\sim$1,500 RGs) and by Hughes et al.\ (2004, 
$\sim$2500 MS, SGB, RG) using the \strom $(m_0, \bmy)$ metallicity 
diagnostic and by S05a using the $(B-V)$ color of $\sim$1,400 RG stars. 

Moreover, $\alpha$-element abundance differences have been detected among 
\omc RGs. They show typical overabundances ($[\alpha/Fe]\sim0.3-0.4$)  
up to iron abundances of \feh\ $\sim$--0.8 (SM00; Cunha et al.\ 2002;
Vanture et al.\ 2002) and evidence of a decrease in the overabundance 
to $[\alpha/Fe]\approx0.1$ in the more metal-rich regime 
(Pancino et al.\ 2002, hereinafter PA02; Gratton, Sneden, \& Carretta 2004). 
The large range of heavy element abundances observed among cluster RGs
was also detected among sub-giant branch (SGB) and main sequence
turn-off (MSTO) stars. In particular, Hilker et al.\ (2004, hereinafter H04) 
and KA06, based on medium-resolution spectra of more than 400 SGB and 
MSTO stars, found a metallicity distribution with three well defined peaks 
around $\feh$ = --1.7, --1.5 and --1.2, together with a handful of metal-rich 
stars with  $\feh\,\sim\,$--0.8. Furthermore, metal-rich stars seem to be 
{\rm CN}-enriched, while metal-poor stars show high {\rm CH} abundances 
(KA06). 
Using the infrared CaT of 250 SGB stars, Sollima et al.\ (2005b, hereinafter S05b) 
found a similar metallicity distribution with four peaks located at 
$\feh$ =--1.7, --1.3, --1.0, and --0.6 dex. The recent metallicity distribution 
based on low-resolution spectroscopy of 442 MSTO and SGB stars by 
Stanford et al.\ (2006a) confirmed the occurrence of a sharp rise 
at low metallicities, $\feh <$ --1.7, and the presence of a metal-rich 
tail up to $\feh\,\approx\,$--0.6 dex. Moreover, Stanford et al.\ (2007), 
using the same spectra, showed that $\sim$16--17\% of the SGB stars 
are enhanced in either {\rm C} or {\rm N} . In order to explain this interesting 
evidence they suggested that these enhancements might be due to primordial 
chemical enrichment either by low (1--3 M/\msun) to intermediate 
(3--8 M/\msun) mass AGB stars, or by massive rotating stars 
(Maeder \& Meynet 2006).

This is the second paper of a series devoted to \strom photometry of GGCs. 
In the first paper we provided new empirical and semi-empirical calibrations 
of the metallicity index $m_1=(\vmb)-(\bmy)$ based on cluster 
RG stars and on new sets of semi-empirical and theoretical color-temperature 
relations (Calamida et al.\ 2007, hereinafter CA07). 

The structure of the paper is as follows. In \S 2 we discuss in detail 
the  multiband \strom ($u,v,b,y$) and near-infrared (NIR, $J,H,K_s$ ) 
images we collected for this experiment. In \S 3 and \S 4
we lay out the data reduction techniques and the calibration strategies 
adopted for the different data sets. Section 5 deals with the selection criteria 
adopted to identify candidate field and cluster RG stars using both optical and 
NIR photometry. In \S 6 we present a new calibration of the \strom $m_1$ metallicity 
index based on the \bmy\ color and compare the new relation with similar relations 
available in the literature. 
In \S 7 we discuss the comparison between spectroscopic and photometric
metallicities of \omc RG stars based on different Metallicity Index Color 
(MIC) relations and different calibrations (empirical, semi-empirical). 
In this section we also address the impact that {\rm CN}-strong stars have on 
photometric metallicity estimates. Section 8 deals with metal abundance 
distributions---in particular, we discuss the occurrence of different 
metallicity regimes---while abundance anomalies ($C,N$, $\alpha$--elements) 
are discussed in \S 9. Finally, in \S 10 we summarize the results of this 
investigation and briefly outline possible future extensions of this medium-band 
photometric experiment.

\section{Observations}\label{obs}
\subsection{\strom data}
A set of 110 {\it uvby\/} \strom images centered on the cluster \omc 
was collected by one of us (LF) over 11 nights between 1999 
March 27 to April 9, with the 1.54m Danish Telescope (ESO, La Silla). 
Weather conditions were good, typically with humidity from 40--60\%,
conditions frequently photometric, and seeing ranging from 1\Sec3
for the $y$ band to 2\Sec3 for the $u$ band.
On clear photometric nights, \omc was observed with exposure times
of 1200s, 480s, 240s, 120s for the {\it uvby\/} bands, while 
on less photometric nights, we used exposure times 
of 2000s, 900s, 600s and 450s. The CCD camera was a
2048$\times$2048 pixel Ford-Loral CCD, with
a pixel scale of 0\Sec39 and a field of view of 13\Min7$\times$13\Min7. 
The CCD has two amplifiers, A and B, and we used A in high-gain mode for our 
observations. The gain and the readout noise (RON) of the CCD
were measured to be $1.3\pm0.01$ e$^-$/ADU
and $8.25\pm0.1$ e$^-$, respectively.

This data set has been supplemented with 30 {\it uvby\/} images of
the SW quadrant of \omc collected by one of us (FG) during four nights in
April 1999 and two nights in June 1999 with the Danish Telescope. 
During these nights a set of 112 HD standard stars was observed in
the {\it uvby\/} bands. These were selected from the catalogs of 
photometric standards by Schuster \& Nissen (1988, hereinafter SN88) and
by Olsen (1993,  hereinafter O93). Table~2 and Table~3 give the log of 
these observations together with the seeing conditions. 
\begin{figure}
\begin{center}
\label{fig1}
\includegraphics[width=8cm, height=7cm]{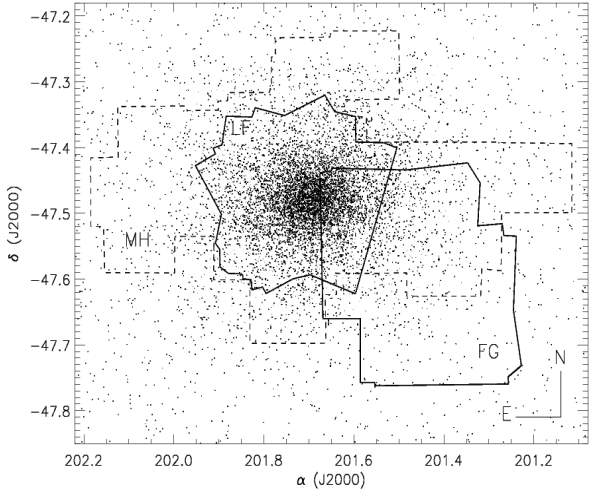}
\vspace{0.15truecm}
\caption{The coverage of the three \strom data sets collected with the 
1.54m Danish Telescope -- L. Freyhammer (LF) central field, F. Grundahl 
(FG) SW fields, and M. Hilker (MH) field, superimposed to \omcp. 
The background is a composite reference image of \omc based on randomly
selected subsamples of stars from the Two Micron All Sky Survey (2MASS)
and the van Leuwen et al.\ (2000) catalog.
}
\end{center}
\end{figure}

Together with the aforementioned images we also acquired 210 {\it vby\/} 
images collected with the same telescope by two of us (MH, TR) in two 
observing runs (1993 and 1995, see Hilker 2000, hereinafter H00; HR00). 
The CCD used for these observations was a Tektronix 1024$\times$1024 chip, with a pixel scale 
of 0\Sec50 and a field of view of 6\Min3$\times$6\Min3.
The log of this data set is given in Table~4 and specifics concerning 
the 33 fields used in this investigation are given in Table~2. 
To obtain a homogeneous photometric catalog, these frames were reanalyzed
adopting the same procedure as was applied to the two other data 
sets. Fig.~1 shows the location of the three different data sets across 
the cluster area.

\subsection{Near-Infrared data}
Near-Infrared $J$, $H$, and $K_s$-band images of \omc were collected in 
two different runs---2001 February 5, 7, and 2002 February 2, 3, 
24, 25---with the NIR camera SOFI at the NTT (ESO, La Silla). 
SOFI can operate at low and high spatial resolution. In the 
former case the pixel scale is 0\Sec292 and the field of view (FoV)  
is 4\Min94$\times$4\Min94, while in the latter the pixel scale is 
0\Sec145 and the FoV is 2\Min47$\times$2\Min47. We observed 
three different fields at low resolution: field $A$ (36$J$, 55$Ks$) 
is located at the cluster center, field {\rm C} (33$J$, 55$Ks$) is located 
$\sim10$\Min5 NW of the center, and field 
$D$ (12$J$, 20$Ks$) is located $\sim10$\Min7 SW of 
the center (see Table~5 and Fig.~2). 
The observing time was split between target observations on \omc and 
reference sky observations. 
For each pointing we acquired a set of $J$ (Detector Integration Time [DIT]=3s, 
Number of DIT [NDIT]=1) and $K_s$ (DIT=3s, NDIT=4) images.
To supplement these data we retrieved from the ESO archive a mosaic of 
nine pointings (11$J$, 11$K_s$; Sollima et al.\ 2004) which 
covers an area of $\sim 13$\min$\times13$\min across the cluster center. 
These images were collected on 2000 January 12, 13, 
using the same equipment in low-resolution mode. 

\begin{figure}
\begin{center}
\label{fig2}
\includegraphics[width=8cm, height=7cm]{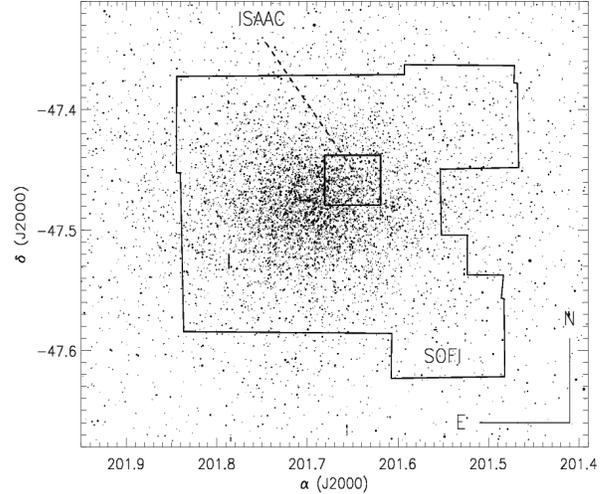}
\caption{Same as Fig.~1, but for a narrower cluster region showing 
the coverage of the two NIR data sets collected with SOFI@NTT and 
with ISAAC@VLT.}
\end{center}
\end{figure}

Moreover, we also retrieved from the ESO archive two more data sets 
collected during the nights 2004 June 3, 4 and 2005 April 2. These data 
sample three different positions: field $\alpha$ (15$J$, low resolution) located at 
the cluster center; field $\beta$ (15$J$, low resolution) located 
$\sim5$\min ~NW of the center; and field $\gamma$ located 
$\sim5$\min ~SE of the center. This last field 
was observed both at low ($\gamma_1$, 12$J$,12$H$,12$Ks$) and high 
($\gamma_2$, 24$Ks$) spatial resolution.  

The above data were further supplemented with a set of NIR images 
from the NIR camera ISAAC at the VLT (ESO, Paranal).  
Its pixel scale is 0\Sec145 and the FoV is 
2\Min5$\times$2\Min5. The images were collected using 
two narrow-band filters centered on $\lambda=1.21\, \mu$m 
($\Delta \lambda=0.018\, \mu$m; 8 images) and 
$\lambda=2.19\, \mu$m ($\Delta \lambda=0.03\, \mu$m; 
24 images). We adopted the same $DIT$ of 6 sec, but  
$NDIT = 13$ and 11 for the $NB\_1.21$ band and the $NB\_2.19$ band, 
respectively. The log of the entire NIR data set is given in 
Table~6, and the area covered by the different SOFI and ISAAC fields 
is shown in Fig.~2.

\section{Data reduction}
\subsection{\strom data}
Flat fields for the central LF field (see Table~2 and Fig.~1) were collected 
for the different filters and CCD settings at dusk and at dawn. 
The Danish Faint Object Spectrograph and Camera (DFOSC) was mounted with its 
File and Shutter Unit (FASU) on the telescope instrument adaptor. 
The 90\,mm \strom filters were installed in the FASU.  The main DFOSC 
instrument with the CCD camera can rotate with respect to the 
FASU around the optical axis, so to average out background gradients in the 
flat-fields---typically introduced by scattered light in the telescope and in
the instrument due to inadequate baffling---we introduced offsets of 90 degrees 
between flat fields in the same bandpass. Because DFOSC is a focal-reducer 
instrument, low-level multiple reflections in the optics are known to cause 
effects such as {\em sky concentration} (Andersen et al.\ 1999). The science 
frames were similarly obtained with rotation offsets. 
For each science image a 2D bias frame was subtracted, then they were 
scaled in level according to prescan and overscan areas, and flat-fielded 
using standard IRAF tasks based on nightly averaged calibration images.
The 90\,mm filters turned out to be affected by the artifact known as 
'{\it filter ghosts}'. The ghosts are located  on the CCD images 
$\sim20$ pixels leftward of saturated stars and typically have the
shape of a flattened doughnut or a fan. The orientation of the
ghost star may slightly change across the CCD field, but on average
it is quite constant. The separation between the saturated star and
its ghost is also rather constant. The orientation of the ghosts in
the CCD images collected with different orientations is rotated 
by the same angle. This evidence indicates that the ghost is not 
caused by double reflected light between the filter and the 
CCD camera, but must occur in the filter itself. 
A similar effect was found by O'Connor et al.\ (1998) in optical 
images collected with the MOMI CCD camera available at the Nordic Optical
Telescope (NOT, La Palma), and also in HST images collected with the 
WFPC2 (Krist 1995).  
The overall effect on our images is of the order of 0.4--1.1\% in 
flux in $b, y$-band images and smaller in $u, v$-band images.
This means that this effect is smaller than the typical 
photometric accuracy along the RGB. However to avoid spurious 
effects in the PSF photometry, we chose no PSF stars located 
in the neighborhood of saturated stars.      

Raw images of the FG data set (see Table~2 and Fig.~1) were pre-processed 
using tasks available in the IRAF data-analysis environment for bias 
subtraction and flat-fielding. To flat-field these data we adopted median 
sky flats collected during the observing nights. The reader interested in the 
details of pre-reduction strategy adopted for the MH images is referred to H00.

The photometry was performed using DAOPHOT$\,${\footnotesize IV}/ALLSTAR and 
ALLFRAME (Stetson 1987; 1991; 1994). We first estimated a
point-spread function (PSF) for each frame from $\approx 100$ bright, 
isolated stars, uniformly distributed on the chip. We assumed a spatially 
varying Moffat function for the PSF. 
We performed preliminary PSF photometry on each image with the task 
ALLSTAR.  Then we used the task DAOMATCH/DAOMASTER (Stetson 1994)
to merge the individual detection lists from the
328 images of the three different data sets into a single global
star catalog referred to a common coordinate system. Twelve 
images were neglected, due to either bad seeing or poor image quality.   
As a reference catalog we adopted $B,V,I$-band photometry for
$\approx$ 270,000 stars of \omc (Stetson 2000). These stars 
are distributed over a field
of about 28\min$\times$28\min\ centered on the cluster and have
been selected for photometric precision 
$\sigma_{B,V,I} \leq$ 0.03 mag\footnote{The catalog of local 
standards can be downloaded from the
{\em Photometric Standard Stars} archive available at the following URL:
http://www1.cadc-ccda.hia-iha.nrc-cnrc.gc.ca/community/STETSON/standards/}.
We then performed simultaneous PSF-fitting photometry over the entire 
data set with the task ALLFRAME.  Aperture corrections were determined
for each frame and applied to the individual stellar magnitudes
before the weighted 
mean magnitudes were determined (Calamida et al.\ 2008b). 
The final merged star catalog includes $\approx 2\times 10^5$ stars 
having measurements in at least two bands and covers a field of about 
23\min$\times$23\min\ centered on \omcp.

\subsection{Near-Infrared data}
The reduction process for the NIR data was the same as for the optical data.
The pre-reduction was performed using standard IRAF procedures, 
and initial photometry for individual images was performed with DAOPHOT/ALLSTAR. 
Then the individual detection lists were merged into a single star catalog and
a new photometric reduction was performed with ALLFRAME.
To improve the photometric accuracy we did not stack individual images, and
the spatially varying PSFs were derived from at least
50 bright, isolated stars uniformly distributed across 
each image.

\section{Photometric calibration}\label{pho}
\subsection{\strom data}
\begin{figure}
\begin{center}
\label{fig3}
\includegraphics[width=8cm, height=7cm]{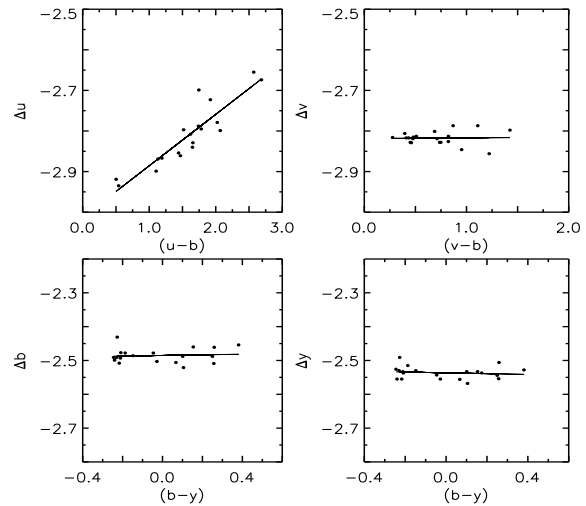}
\vspace{0.15truecm}
\caption{Calibration curves for the reference night June 6.}
\end{center}
\end{figure}

Standard stars for the \strom photometry were selected from the lists of Henry
Draper stars published by SN88 and O93; these were
observed during four nights in April 1999 and two nights in 
June 1999 (see Grundahl, Stetson, \& Andersen  2002 for more 
details).  Table~7 and 8 show the observation logs for these 
stars. We applied the same pre-reduction strategy and performed
aperture photometry on the standard frames, ending up 
with a list of $\approx$ 20 HD stars per night.
Extinction coefficients were estimated from observations
of HD stars obtained at different airmass values (see Table~9). 
Coefficients estimated for three out of the four nights in April are 
mutually consistent, but the values obtained for June 6 are 
slightly larger (see Table~10).  
We estimated independent color transformations for each night and 
thus obtained a set of six calibration curves. These 
agree quite well with each other and we finally selected the best 
photometric night (June 6) to perform an absolute calibration of 
the scientific frames collected during this night.
The calibration curves for the reference night are (see also Fig.~3):

$u= u_i - (3.012\pm0.015) + (0.127\pm0.010)\times(u_i-b_i)$,

$v= v_i - (2.820\pm0.008) + (0.002\pm0.016)\times(v_i-b_i)$,

$b= b_i - (2.488\pm0.012)$,

$y= y_i - (2.736\pm0.010)$.

\noindent where $i$ stands for the $instrumental$ magnitude.

On the basis of this calibration we defined a set of secondary standard 
stars, by selecting both according to photometric precision ($\sigma_{u,v,b,y} \leq$ 
0.05 mag) and according to the ''separation index'' 
({\tt $sep_{v}$ > 5})\footnote{The ''separation index'' 
quantifies the degree of crowding (Stetson et al.\ 2003). The current 
$\tt sep$ value allow us select stars that have required corrections
smaller than $\sim$1\% for light 
contributed by known neighbours.} in the calibrated star catalog of the 
reference night. We ended up with a sample of $\sim$5000 stars having
measurements in all the \strom bands.  We extended the photometric
system of the reference data to each overlapping field by an iterative
procedure.

In estimating the mean calibrated magnitudes, we neglected the $b$ 
and $y$ frames of the LF data set because the estimate of the 
aperture correction was hampered by the severe stellar crowding.  
For two pointings of the MH data set, namely fields $q$ and $r$ 
(see Fig.~1 of H00), the number of local standards was too small 
to provide a robust estimate of the calibration curves and they 
have not been included in the final catalog.

\begin{figure}
\begin{center}
\label{fig4}
\includegraphics[width=8cm, height=7cm]{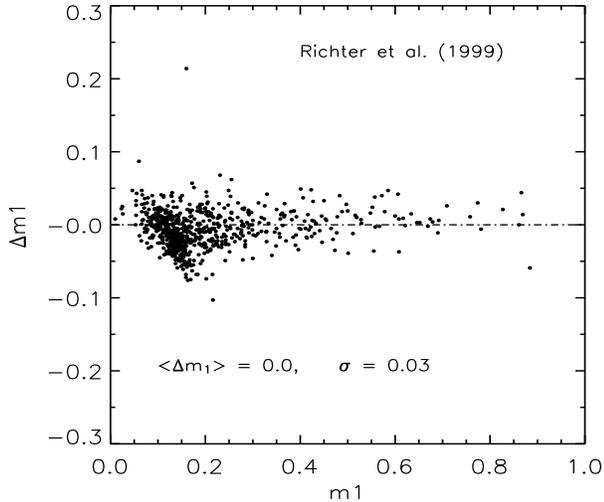}
\caption{Comparison between the $m_1$ index of \omc stars 
measured by HR00 and current photometry.}
\end{center}
\end{figure}

A slightly different approach was applied to the MH 1993 data. This data set 
included many non-photometric nights (see Table~4), and there were only a few
secondary standards in this region of \omcp. Therefore, the estimate 
of the relative zero-points between the different frames was difficult. 
Thus we used the final calibrated star catalog based on the other observations as a new set of local 
standard stars to individually calibrate each frame of the 1993 data set. 
However, fields 3, 4 and 8 (see Fig.~1 of HR00) were not included because 
the number of secondary standards was still too small in this cluster region.
The typical accuracy of the absolute zero-point calibration is 
$\sim$ 0.02 mag for the $u,v$-band data and $\sim$ 0.015 mag for 
the $b, y$-band data.

We compared our calibration with the absolute calibrations provided by 
Richter et al.\ (1999), by Mukherjee et al.\ (1992) and by Hughes \& 
Wallerstein (2000). Fig.~4 shows the comparison between the $m_1$ index 
based on the current photometry and the $m_1$ index based on the Richter et al.\ 
photometry. The mean difference is $\Delta(m_1)\sim$ 0.00 with a dispersion 
of $\sigma\sim$ 0.03 mag. The mean differences between our and the other two 
studies are $\Delta(m_1)\sim$ 0.01, with $\sigma\sim$ 0.06 based on 139 stars 
in common with the Mukherjee et al. catalog, and $\Delta(m_1)\sim$ -0.02, 
with $\sigma\sim$ 0.05, based on 228 stars in common with the Hughes \& Wallerstein catalog. 

\begin{figure*}
\begin{center}
\label{fig5}
\hspace*{-1.5cm}
\includegraphics[height=10cm,width=20cm]{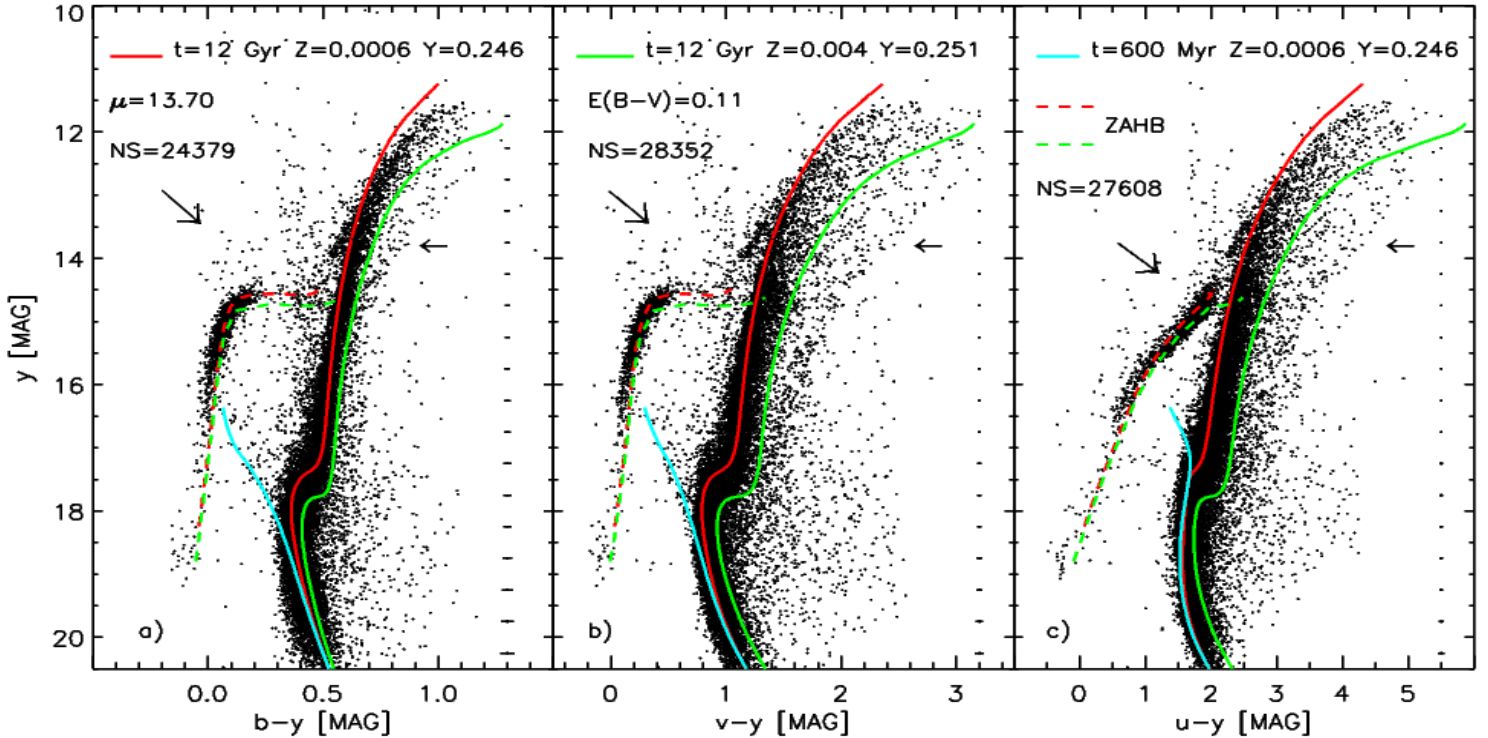}
\caption{From left to right a) $y$, \bmy;  b) $y$, \vmy;  
c) $y$, \umy~ \strom CMDs of \omcp. Stars 
were selected according to photometric accuracy---$\sigma_{u,v} \le$ 0.05 mag, 
$\sigma_{b,y} \le$0.03 mag---and separation---$sep_v \ge 2$, $sep_{b,y} \ge 3$. 
The red and the green solid lines represent two cluster isochrones at fixed 
age ($t=12$ Gyr) and different chemical compositions (see labeled values). 
The dashed lines show two ZAHBs for the same chemical compositions and a 
progenitor mass of 0.8 $M/M_\odot$. 
The turquoise solid line shows a young ($t=600$ Myr), metal-poor cluster 
isochrone.  The true distance modulus 
($\mu$) 
and the reddening adopted to 
overplot the theoretical predictions are also labeled. 
Error bars plotted on the right side show the intrinsic photometric uncertainties 
in color at different $y$ magnitudes. The horizontal arrows mark the location 
of the \omtre~ metal-rich RGB, while the other arrows represent the reddening 
vector.}
\end{center}
\end{figure*}

The off-center field covered by the Mukherjee et al.\ (1992) {\it uvby}-band 
photometry does not overlap with fields for which $u$-band images are  
available in our catalog. On the other hand, an independent check on the 
accuracy of our $u$-band calibration has been performed comparing the 
location of hot horizontal branch (HB) stars ($T_{eff} \ge 8500 K$) in the 
$[c]=[(u-b)-(b-v)]-0.2\times(b-y)$ vs \umy\ plane with similar stars in  
other globular clusters (Calamida et al.\ 2005, hereinafter CA05).
Finally, we estimated the weighted mean magnitudes between the different
data sets, ending up with a merged catalog that includes $\approx$ 185,000 stars,
with an accuracy of $\sim$ 0.03 mag at y$\sim$ 19 mag. To our knowledge this  
is the largest \strom photometric catalog ever collected for a GC.

The final catalog has been filtered by photometric accuracy and stars have 
been plotted in the $y$, \bmy\ (left panel), $y$, \vmy\ (middle) 
and $y$, \umy\ (right) CMDs (Fig.~5). In order to verify the current 
absolute calibrations we performed a detailed comparison with theoretical 
predictions. In particular, we adopted the estimated true \omc distance modulus  
provided by Del Principe et al.\ (2006), $\mu=13.70\pm0.06$, and a mean 
reddening value of $E(B-V)=0.11\pm0.02$ (CA05). The reader interested in 
a more detailed discussion of the \omc distance is referred to 
Bono et al.\ (2008). The extinction coefficients for the \strom colors 
were estimated from the Cardelli et al.\ (1989) reddening law: 
$E(m_1)= -0.3\times E(b-y)$, $E(b-y)= 0.70\times  E(B-V)$, 
$E(v-y)= 1.33\times E(B-V)$ and $E(u-y)= 1.84\times E(B-V)$.
The solid lines in Fig.~5 show two cluster isochrones at fixed age 
($t = 12$ Gyr) and different chemical compositions, $Z=0.006, 
Y=0.246$ and $Z=0.004, Y=0.251$, while the dashed lines represent the 
corresponding Zero Age Horizontal Branch (ZAHB) models. 
The evolutionary models, the bolometric corrections and the 
color-temperature relations adopted in this investigation were 
discussed by Cassisi et al.\ (2004) and by Pietrinferni et al.\ (2006). 
The stellar models and the color transformations were computed by 
assuming an $\alpha-$element enhancement of 
$[\alpha/Fe]=+0.4$\footnote{The adopted set of models can be 
downloaded   
from the BaSTI archive available at the following URL: 
{\tt http://www.oa-teramo.inaf.it/BASTI}}.
Data plotted in Fig.~5 show that theory and observations agree
within the errors over the entire magnitude range. 
In particular, the two old ($t=12$ Gyr) isochrones bracket the metallicity 
spread of the bulk of \omc RG stars ($-1.8 \lesssim \feh \lesssim -1.0$). 
(This result does not apply to the anomalous metal-rich RGB, $\omega_3$,  
(Lee et al.\ 1999; PA00; Freyhammer et al.\ 2005) marked 
with a horizontal leftward arrow in the figure. A detailed 
fit to this small sub-population is not a goal of the current investigation.)
On the other hand, the \vmy\ and the \umy\ colors of some RGs are redder 
than those predicted by models. 
This scatter towards redder colors might be due to differential reddening 
or to {\rm CN/CH} enhancements.  Such enhancements typically produce
redder  
\vmy\ / \umy\ colors (Grundahl et al.\ 2002; CA07) than for chemically
normal stars. Note that in a recent 
investigation Stanford et al.\ (2007) suggested that $\approx$ 16--17\% 
of \omc SGB stars are either {\rm C} or {\rm N} enhanced. 

The Blue Straggler (BS) sequence can be easily identified in the 
$y$, \bmy\ and in the $y$, \vmy\ CMDs, but they can barely be identified 
in the $y$, \umy\ CMD. This effect is due to the fact that the 
$u$ and the $y$ magnitudes of main-sequence (MS) structures ($\log \sim 4.75$), 
at fixed metal content ($Z\sim0.0006$), show similar decreases 
(1.42 vs 1.35 mag) when moving from an effective temperature of 
$\approx 10,000$ K to $\approx 7,000$ K (see the turquoise 
600 Myr isochrone plotted in Fig.~5). In comparison, the 
$b$ and the $v$ magnitudes decrease, in the same temperature range, 
by 1.58 and 1.74 mag.  

The HB is well populated ($\approx$ 2550 stars) in all the CMDs, 
and among them we find $\approx$ 270 are Extreme Horizontal Branch (EHB) stars, 
i.e., HB stars with $y\ge 17.8$ and \umy$\sim 0$. It is noteworthy that 
in the same cluster area we identified $\approx 2,900$ HB stars 
in ground-based images collected with the Wide Field Imager 
(WFI@2.2m ESO/MPG, La Silla) and in images collected with the 
Advanced Camera for Surveys (ACS, Hubble Space Telescope [HST];
Castellani et al.\ 2007, hereinafter CAS07). The combined 
catalog (WFI$+$ACS) can be considered complete in the magnitude 
range ($14.5 \lesssim y \lesssim 18$). 
These star counts indicate that the current \strom images have allowed us 
to identify $\sim$ 90\% of the HB stars measured in the WFI--ACS images. 
The \omc CMDs also show that our photometric catalog is contaminated by 
field stars, see e.g.\ the vertical plume of stars located at $y\le 14.5$ 
and \bmy$\sim0.4$, \vmy$\sim0.9$,  \umy$\sim1.7$ mag. In section 5, we 
briefly describe and adopt the procedure devised by CAS07 and by CA07
to properly identify candidate field and cluster stars.

\subsection{Near-Infrared data}

The final calibrated NIR star catalog includes $\sim$150,000 stars with photometric
precision better than 0.1 mag, i.e., objects with at least ten $\sigma$ detection 
significance. Panel a) of Fig.~6 shows the $K_s, J-K_s$ 
CMD selected according to $\sigma_{J,K_s} \le$0.045 mag and 
{\tt $sep_{K_s}$ > 10}~mag.  The current NIR 
photometry has very good accuracy well below the TO region. To our 
knowledge this is the largest IR CMD ever collected for a globular cluster. 
This notwithstanding, we have only marginally detected EHB 
stars ($K_s \gtrsim 17.5$) since they are very faint in the NIR bands.     
The cluster isochrones and the ZAHBs plotted in this figure are the same 
as in Fig.~5. We also adopted the same distance and reddening, while 
for the extinction coefficients we have used $A_J = 0.282\times A_V$ and 
$A_{K_s} = 0.116\times A_V$, according to the Cardelli et al.\ reddening 
law. Theoretical models were transformed from the Bessell \& Brett (1988) 
NIR photometric system to the 2MASS photometric system using the 
transformations provided by Carpenter (2001). 
The cluster isochrones bracket, within the errors, the bulk 
of evolved and main sequence stars, while the ZAHBs appear to be, at fixed
color, slightly brighter than hot HB stars. 

\begin{figure*}
\begin{center}
\label{fig6}
\hspace*{-1.5cm}
\includegraphics[height=10cm,width=20cm]{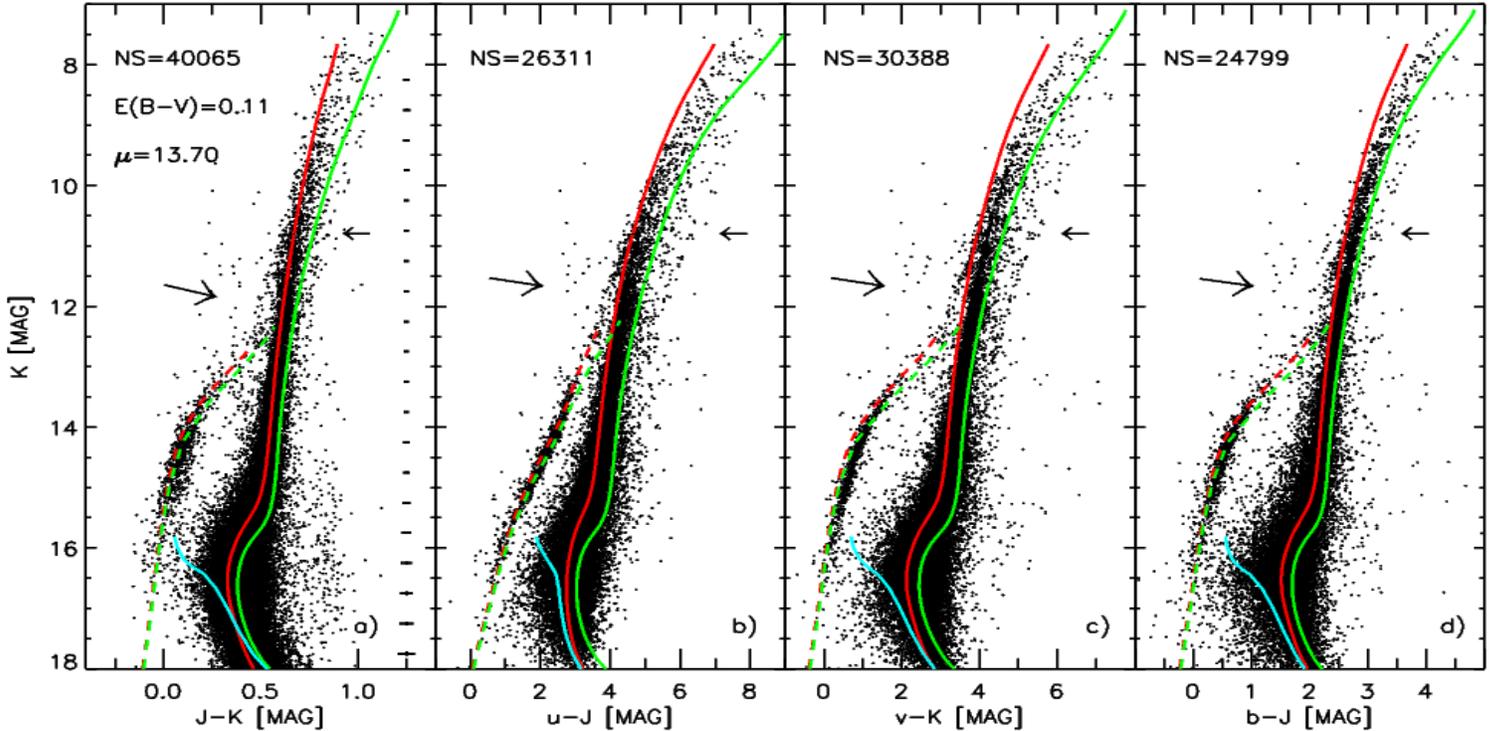}
\caption{\label{fig:fig6} From left to right NIR a) $K$, \jmk, 
and \strom--NIR b) $K$, \umj; c) $K$, $v-K$; d) $K$, \bmj CMDs 
of \omcp. Stars are selected according to photometric accuracy 
$\sigma_{J,K,u,v,b}\le 0.05$ mag and separation {\tt $sep_{K_s}$ > 5}. 
Cluster isochrones, ZAHBs, true distance and reddening are the same 
as in Fig.~5. The error bars in the left panel show the mean intrinsic 
error in color at different $K$ magnitudes. The small 
horizontal arrows mark the location of the \omtre~ metal-rich RGB, 
while the other arrows display the reddening vector.}
\end{center}
\end{figure*}

\section{Selection of RG stars}\label{sele}
To study the \omc RG metallicity distribution, we selected only stars 
with magnitudes $y \lesssim $ 16.5 mag, ending up with a sample of 
$\approx$ 13,000 stars. However, in order to avoid subtle systematic 
uncertainties in the metallicity estimates, this sample needs 
to be purged of the contamination of field stars. To do 
this we used the optical-NIR color-color planes suggested 
by CAS07 and by CA07.  
The \strom photometry was cross-correlated with our NIR photometry in the 
central cluster regions (FOV $\approx 13\arcmin\times13\arcmin$, see Fig.~2), 
while for the external regions we adopted the Two Micron All Sky Survey 
(2MASS) NIR catalog (see \S 2.2). The 2MASS catalog has an astrometric 
precision of $\sim$ 100 mas, and the magnitude limit ranges from 
$\sim 14.3$ for the $K$-band to $\sim 15.8$ for the $J$-band. 
In \omc these magnitude limits fall near the base of the RGB and the TO.
The positional cross-correlation was performed with DAOMASTER, and we 
ended up with a sample of $\sim$ 64,000 stars having measurements in at 
least three of the four \strom bands {\it and\/} at least two of the 
three NIR bands. 

Fig.~6 shows representative optical-NIR CMDs, namely $K$, \umj\ (panel b), 
$K$, \vmk\ (panel c), and $K$, \bmj\ (panel d). The temperature 
sensitivity of optical-NIR colors is quite large, and indeed the 
\umj~ color increases by eight magnitudes when moving from hot HB stars 
(\umj$\sim 0$) to cool giant stars (\umj$\ge 8$) near the 
RGB tip. The same applies to the \vmk~ color, where the aforementioned 
star groups span a range of $\approx 7$ mag.   
In particular, the discrete nature of \omc RGBs (S05a) and 
the split between AGB and RGB stars can be easily identified in the three 
optical-NIR CMDs. The red and green lines display the same cluster 
isochrone as in Fig.~5, and they correspond to the same distance 
modulus and extinction coefficients. Data plotted in Fig.~6 show that, within 
the current uncertainties, theory and observations agree quite well. It is worth 
noting that BSs in the $K$, \bmj\ CMD (panel d) form a spur of blue stars 
well separated from the bulk of the old population at $K \sim 16.5$,
\bmj$\sim$0.9.   A more detailed comparison between theory and optical-NIR 
data will be discussed in a future paper.  

In what follows, we have adopted the \vmk vs \umj color-color plane to fully exploit
the temperature sensitivity of optical-NIR colors in distinguishing cluster from
field stars.


\begin{figure}
\label{fig7}
\includegraphics[width=8.5cm, height=13cm]{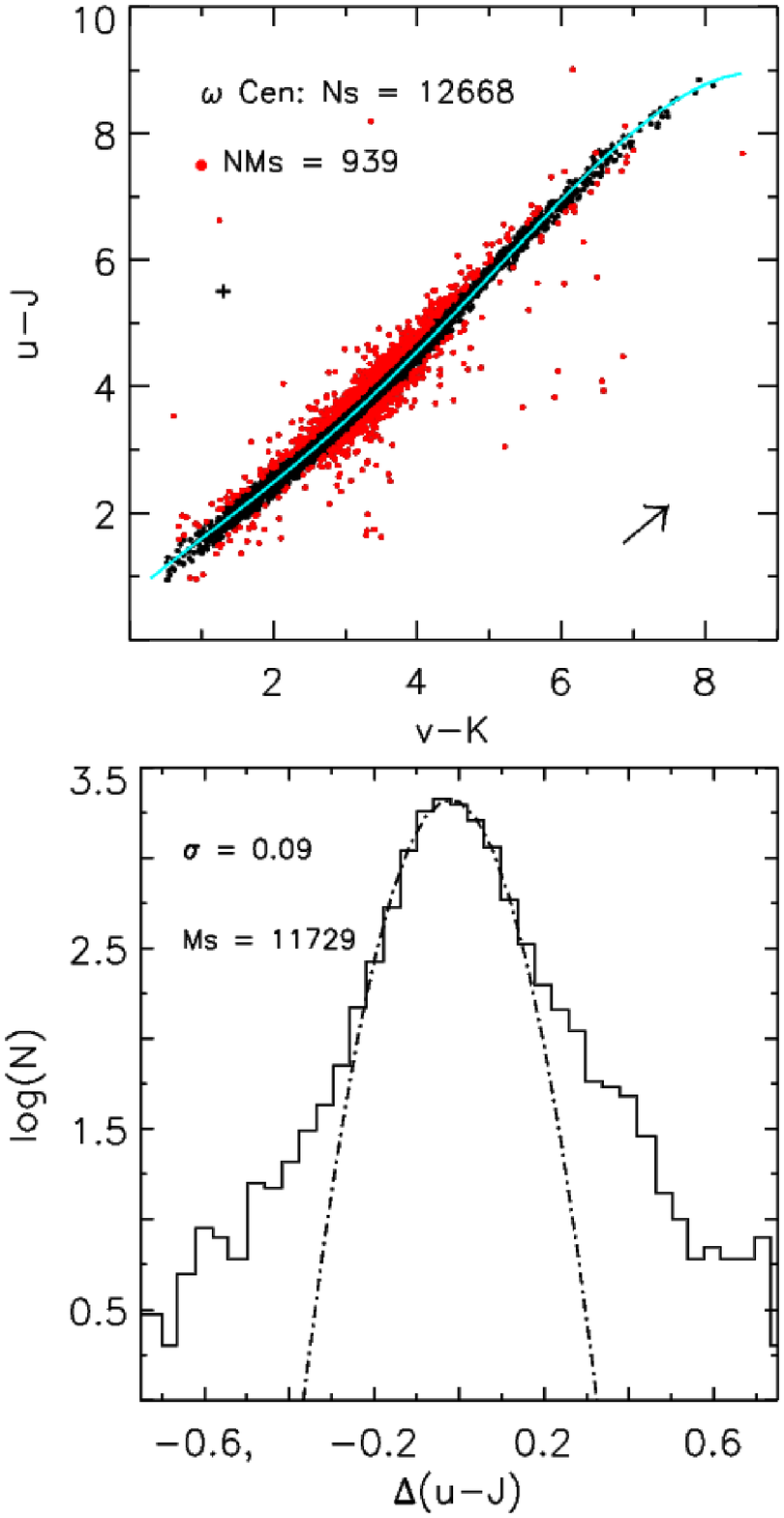}
\caption{\label{fig:fig7}\small{Top: \omc RG stars plotted in the
$(\umj, \, \vmk)$ plane. The error bars display the mean photometric error
in the adopted colors. Red dots are candidate field stars (Non-Member stars, 
NMs = 939). 
The solid turquoise line is the fitted cluster fiducial sequence.
Bottom: logarithmic distribution of the difference between the \umj\ color
of individual stars and the \umj\ color of the fiducial cluster sequence. 
The dashed-dotted line displays the Gaussian function that fits the 
color difference distribution. Objects with $\Delta (\umj) \le 2.5\times
\sigma_{\umj}$ were considered candidate \omc RG stars (Member stars, 
Ms = 11,729).}}
\end{figure}

We first selected RG stars with measurements in the four \strom 
bands ($u,v,b,y$) and two NIR bands ($J,K$), ending up with a sample of 
more than 12,000 stars (see top panel of Fig.~7).  
To define the fiducial cluster sequence in the 
$\umj, \, \vmk$ plane we selected the stars at distances 
from 0\Min5 to 1\min\ from the cluster center (see black 
dots in the top panel of Fig.~7). In this plane the distribution 
of the cluster RG sequence is not linear, so we adopted a 
fourth-order polynomial to represent the RGB. 
\footnote{CA07 found that in the $\umj$ vs $\bmh$ color-color
plane, RG stars in typical GCs obey linear trends. 
The nonlinear trend we have found for \omc RG stars here is due partly to 
the adoption of a different optical-NIR color-color plane and partly 
to the metallicity dispersion among \omc stars.}
Then, we estimated the difference in color between individual RG stars 
and the \umj\ color of the fiducial line at the same \vmk\ color. The bottom panel of Fig.~7 shows 
the logarithmic distribution of the color excess $\Delta (\umj)$ for the entire 
sample. We fitted the distribution with a Gaussian function and we 
considered only those stars with $\Delta(\umj) \le 2.5\times \sigma_{\umj}$ as 
{\em bona fide} \omc RG stars. The red dots in the top panel of Fig.~7 mark 
the candidate field stars after this selection. 

\begin{figure}
\label{fig8}
\includegraphics[width=8.5cm, height=13cm]{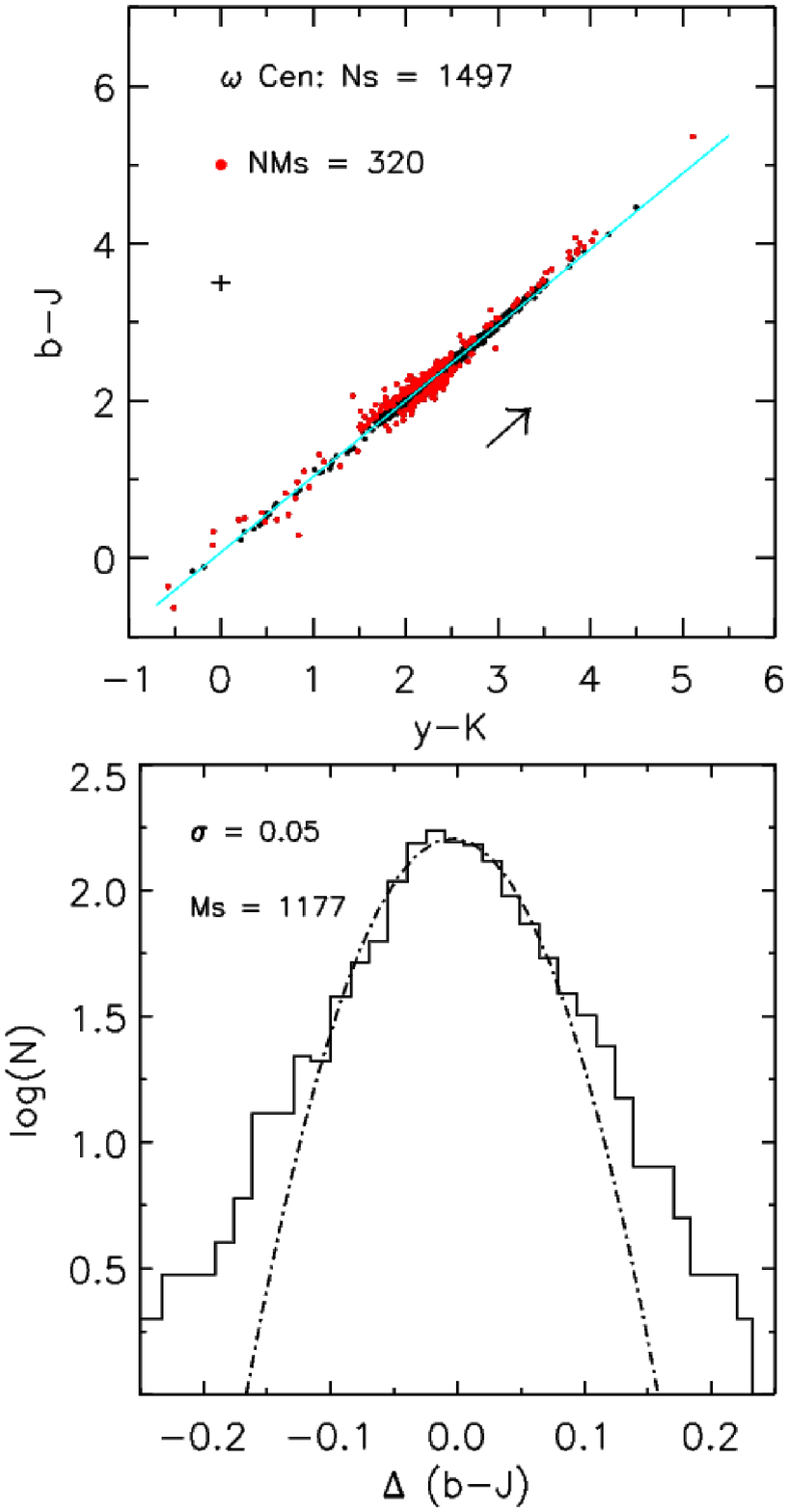}
\caption{\label{fig:fig8}\small{Top: \omc RG stars plotted in the
$(\bmj,\,\ymk)$ plane. The error bars show the mean photometric
error in the adopted colors. Red dots are candidate field stars (NMs = 320).
The solid turquoise line is the fitted cluster fiducial sequence.
Bottom: logarithmic distribution of the difference between 
the \bmj\ color of individual stars and the \bmj\ color of the 
fiducial cluster sequence. The dashed--dotted line displays the 
Gaussian function that fits the distribution of the difference 
in color. Objects with $\Delta (\bmj) \le 1.5\times \sigma_{\bmj}$ 
were considered candidate \omc RG stars (Ms = 1,177).}}
\end{figure}

Subsequently, we selected only stars from the MH data set with 
measurements in the \strom $v,b,y$ bands (1497 RGs) {\it and\/} in 
the $J,K$ NIR bands that were not included in the previous selection.
In this case, we adopted the $\bmj, \, \ymk$ plane to clean the RG sample 
of field star contamination. Fig.~8 shows the selected stars plotted in 
this plane, and the logarithmic distribution (bottom panel) of the color 
excess $\Delta (\bmj)$. In fitting the distribution with a Gaussian function, 
we considered only those stars having $\Delta ({\bmj}) \le 1.5\times \sigma_{\bmj}$ 
as {\em bona fide} \omc RG stars. The red dots in the top 
panel of Fig.~8 mark the candidate field stars in this sample.
Since we applied different selection criteria to the different samples,
we ended up with a total of $\sim 12,900$ RGs with a measurement in at least 
three \strom bands and in two NIR bands. The original sample was thus reduced 
by roughly 10\%. 

Subsequently, in order to improve the selection of candidate field stars
we also re-identified a subsample of our optical catalog in the
proper motion catalog of van Leeuwen et al.\ (2000, hereinafter LE00).
The average proper-motion precisions in this catalog
range from 0.1 mas yr$^{-1}$ to 0.65 mas yr$^{-1}$ from the 
brightest to the faintest stars ($V\sim20$ mag). The average positional 
errors are 14 mas. The cross identification was performed in
several steps: (1) the IRAF/IMMATCH package was used to establish
a preliminary spatial transformation from the \strom catalog
CCD coordinates to the reference catalog equatorial (J2000.0)
system for a subsample of matched stars; (2) the full \strom
catalog was transformed onto and matched with the reference
catalog on the basis of positional coincidence and initially also on apparent
stellar brightness; (3) the previous steps were reiterated
two or three times until the ultimate transformation permitted a near-complete
matching, and then (4) the final, matched sample of common
stars excluded entries separated by more than 0.56".
This provided us with reliable equatorial coordinates and proper
motions for matching stars in the \strom data set.

Then, we performed a selection by proper motion. In particular, 
we considered as cluster members those stars with membership probabilities 
(see LE00) higher than 65\%. This additional 
selection further decreased the sample of candidate RG members by $\approx$ 2\%. 
The final catalog includes $12,700$ probable \omc RG stars.

\begin{figure*}
\label{fig9}
\begin{center}
\includegraphics[height=0.5\textheight,width=0.7\textwidth]{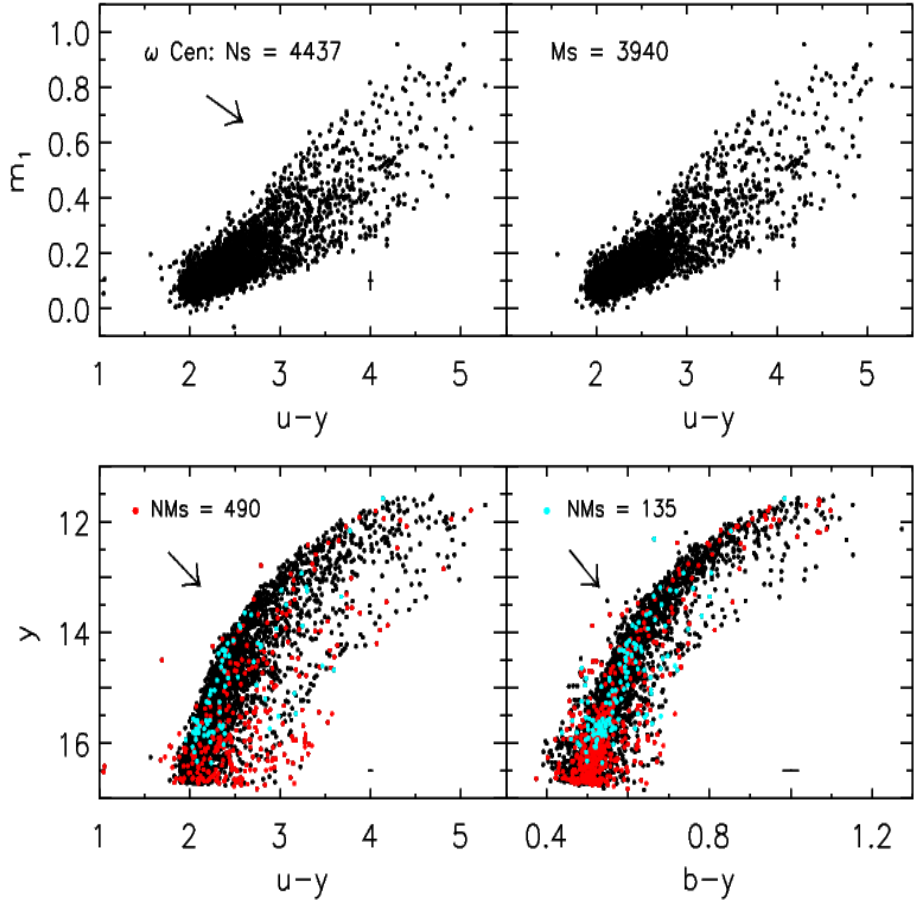}
\caption{\label{fig9}\small{Top: \omc RG stars selected in photometric
error ($\sigma_{v,b,y} \le$ 0.015 mag) and separation ($sep_{v,b,y} \ge$ 3) 
and plotted in the $m_1,\, \umy$ plane (left). The right panel shows candidate 
\omc RG stars (Ms = 3,940) plotted in the same plane. Candidate cluster stars 
were selected according to optical--NIR colors 
($\Delta (\umj) \le 2.5\times \sigma_{\umj}$ and $\Delta (\bmj) \le 1.5\times
\sigma_{\bmj}$) and membership probability higher than 65\%. 
Bottom: \strom CMDs $y$, \umy\ (left) and $y$, \bmy\ (right)
for cluster and field star candidates. The black dots mark candidate 
\omc RG stars, while red dots mark candidate field stars selected according to the 
optical--NIR color--color selection (NMs = 490). The turquoise dots mark probable 
field stars according to the proper motion selection (NMs = 135). The error 
bars account for uncertainties in intrinsic photometric errors. The arrows 
show the reddening vector.}}
\end{center}
\end{figure*}

\begin{figure}
\label{fig10}
\begin{center}
\includegraphics[height=0.3\textheight,width=0.4\textwidth]{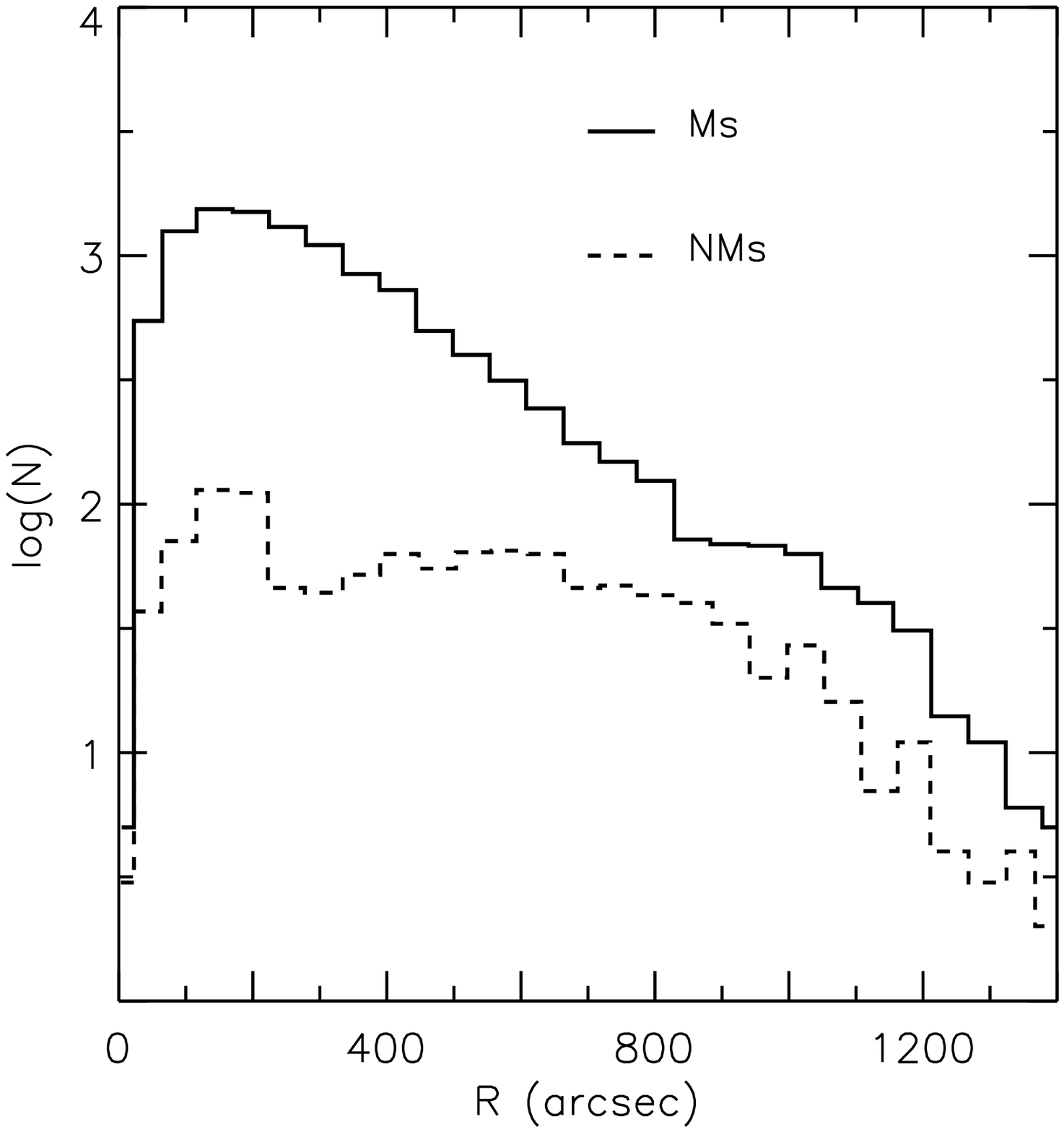}
\caption{\small{Radial distribution of candidate \omc RG 
stars 
(solid line) and of field stars (dashed line). The latter sample includes 
stars selected according to the optical--NIR color--color planes and to the 
proper motions.}
}
\end{center}
\end{figure}

To test the plausibility of the procedures adopted
to disentangle \omc members from field stars, the top panels of Fig.~9 show from 
left to right the distribution in the $m_1,\umy$ plane of 
the original RG sample selected by photometric precision ($\sigma_{v,b,y} \le$ 0.015 mag
and $sep_{v,b,y} \ge$ 3) and of the sample after the selection in 
the optical-NIR color-color plane. Fig.~9 shows that \omc RGs do not follow a 
well-defined sequence on the $m_1,\umy$ plane, as the other GCs do (see CA07).
The RG stars in \omc show a wide spread in color that cannot be explained as 
photometric errors or differential reddening (see the error bars and the reddening 
vector plotted in Fig.~9). The bottom panels of Fig.~9 show the same RG sample 
(black dots) plotted in the $y$, \umy\ (left) and in the $y$, \bmy\ (right) CMDs;
overplotted are the candidate field stars after the selection in the optical-NIR 
color-color plane (red dots), and after the selection by proper motion 
(turquoise dots).
It is noteworthy that most of the field stars have magnitudes and colors that are 
very similar to the probable cluster RG stars.

As a final validation of the two selection procedures, we compared the radial 
distribution of candidate \omc RGs and field stars. Fig.~10 shows
the two logarithmic distributions and the flat trend of field stars 
(dashed line) is quite evident when compared with the steeper and more 
centrally concentrated distribution of likely \omc stars (solid line).
It was already noted by CA07 that the mild decrease in the number of apparent field 
stars in the outskirt of the cluster suggests that our color-color selection 
is very conservative, and some real cluster members have probably been erroneously 
rejected. However, in these photometric selections it is more important
to keep the probable nonmembers out than to keep probable members 
in (ATT00).

\section{A new calibration of the $\muno$, $\bmy$ metallicity relation}\label{rela}
In CA07 we provided new empirical and semi-empirical calibrations of the $m_1$ 
metallicity index based only on cluster RG stars. These new 
MIC relations were then validated on the basis of
GC and field RGs with known spectroscopic iron abundances in the 
Zinn \& West (1984) metallicity scale, and were found to provide metallicity estimates 
with an accuracy of the order of 0.2 dex. The main difference between the CA07 
MIC relations and similar relations available in the literature is that 
the former adopt the \umy\ and \vmy\ colors instead of \bmy. The main advantage 
of the \umy\ and \vmy\ colors is the stronger temperature sensitivity, and 
the MIC relations show a linear and well-defined slope in the 
$m_1,\umy / \vmy$ planes. However, the \umy\ and the \vmy\ colors also 
have a drawback: they are affected by carbon (C) and/or nitrogen (N) 
enhancements (CA07). 
To investigate the different behavior of the \umy, \vmy, and \bmy\ 
colors as a function of the {\rm CH} and the {\rm CN} abundances, we also performed a 
metallicity calibration based on the \bmy\ color.
We found that this MIC relation shows a nonlinear trend over the typical 
color range of RG stars, namely $0.42<(\bmy)_0<1.05$ (see Fig.~2 of 
Calamida et al.\ 2007b, hereinafter CA07b). 
Therefore, we included a quadratic color term to characterize the calibration 
(see also HR00 and CA07b). 
Specifically, we estimated a semi-empirical calibration based on the \bmy\ 
color of the theoretical evolutionary models by Pietrinferni et al.\ (2006)
for an $\alpha$-enhanced ($\afe=0.4$) mixture. Theoretical predictions were 
transformed to the observational plane with the semi-empirical Color-Temperature Relations (CTRs) 
provided by Clem et al.\ (2004). We adopted the procedure described in CA07 
and applied a multilinear regression fit to estimate the coefficients of 
the MIC relation:

\begin{equation}
m_{0} = \alpha + \beta\,\feh + \gamma\, CI_0 +
\delta\,\feh\, CI_0 + \epsilon\, CI_0^2 + \zeta\,\feh\, CI_0^2 
\end{equation}

where $m_0$ = $m_{10}$ and $CI_0$ are the unreddened color indices. 
Similar relations were also derived for the reddening-free 
$[m]=m_1 + 0.3\times(b-y)$ index. 
Together with the calibration of semi-empirical MIC relations we 
performed an independent empirical calibration using the observed colors 
of RG stars in selected GCs, M92, M13, NGC$\,$1851, NGC$\,$104, for which 
accurate \strom photometry is available (Grundahl et al.\ 1998; CA07). 
These GCs cover a broad range in metal abundances ($-2.2<\feh<-0.7$) and
are minimally affected by reddening ($E(\bmv) \le 0.04$).
The coefficients of the fit of both semi-empirical and empirical MIC relations 
are listed in Table~11.
In order to validate the new MIC relations we decided to apply 
them to nine GCs for which accurate \strom photometry, absolute 
calibration, and sizable samples of RG stars are available. They are M92, 
NGC$\,$6397, M13, NGC$\,$6752, NGC$\,$1851, NGC$\,$288, NGC$\,$362, M71 and
NGC$\,$104. Table~12 lists spectroscopic and photometric metallicities for 
these clusters. Note that five out of the nine GCs 
had been adopted to calibrate the empirical MIC relations. We find a good 
agreement for both the empirical and the semi-empirical MIC relations. The 
comparison between the mean photometric metallicity estimates and the spectroscopic 
measurements indicates that they agree with each other within $1\sigma$ uncertainties. 
It is noteworthy that the accuracy of photometric metallicity estimates 
applies not only to metal-intermediate clusters with low reddening corrections 
(NGC$\,$288, $E(B-V)\sim0.02$), but also to more metal-rich clusters 
with high reddening (M71, $E(B-V)\sim0.30$, Harris 1996).

To further constrain the calibration of the new MIC relations we decided to 
compare them with $m_0, (\bmy)_0$ metallicity calibrations available in the 
literature. We adopted the field RG sample by Anthony-Twarog \& Twarog (1994, 
hereinafter ATT94) and Anthony-Twarog \& Twarog (1998, hereinafter ATT98), already 
used by CA07. The interested reader is referred to this paper for more details 
concerning the selection of the RG sample.

We compared the photometric metallicities of 59 out of the 79 stars
for which ATT94 provided an estimate. The spectroscopic and the photometric 
metallicity distributions are in fair agreement. 
The mean differences are $0.02\pm0.03$, with a dispersion of $\sigma$ = 0.22 dex
(ATT94) and $0.13\pm0.03$, with a dispersion of $\sigma$ = 0.20 dex (H00). The small
difference between the current errors and the original ones derived by ATT94 and H00
is only due to the different selection criteria. The mean differences using the same
59 stars and current empirical and semi-empirical MIC relations are $-0.12\pm0.02$
and $-0.15\pm0.02$ dex, with a dispersion of $\sigma$ = 0.15 and of $\sigma$=0.18 dex,
respectively. 
Fig.~11 shows the comparison between the spectroscopic metallicity distribution 
(dashed line) and the metallicity distribution obtained with the empirical 
MIC relation based on the ($m_0,(\bmy)_0$) calibration by ATT94. The middle panel 
shows the comparison with the metallicity distribution based on our empirical 
and semi-empirical ($m_0, (\bmy)_0$) MIC relations for the complete sample of
79 RG stars. The bottom panel shows the comparison with the metallicity distribution 
based on the empirical MIC relation ($m_0,(\bmy)_0$) provided by H00 for 73 out of 
the 79 RG stars with $-2.3<\feh<-0.4$.
Data plotted in the top and in the middle panel of Fig.~11 indicate that spectroscopic 
and photometric metallicity distributions agree quite well. The metallicity distribution
based on the H00 calibration appears slightly more metal-rich ($\sim$ 0.13 dex). 
These data show that metal-abundance estimates for the ATT98 
sample based on the ($m_0, (\bmy)_0$) MIC relations are shifted by  
$\sim$ -0.14 dex, as already found by CA07 for empirical and semi-empirical 
($m_0, (\vmy)_0$; $m_0, (\umy)_0$) relations. They suggested that this 
difference might be due the different approach (cluster vs field RGs) in 
the calibration of the $m_1$ index. 
In spite of this systematic difference, the intrinsic dispersion of the difference
between photometric and spectroscopic metallicities is smaller than 0.2 dex, and it
is mainly due to photometric, reddening, and spectroscopic errors
(see Fig.~2 of CA07b). 

\begin{figure}
\label{fig11}
\begin{center}
\includegraphics[height=0.5\textheight,width=0.5\textwidth]{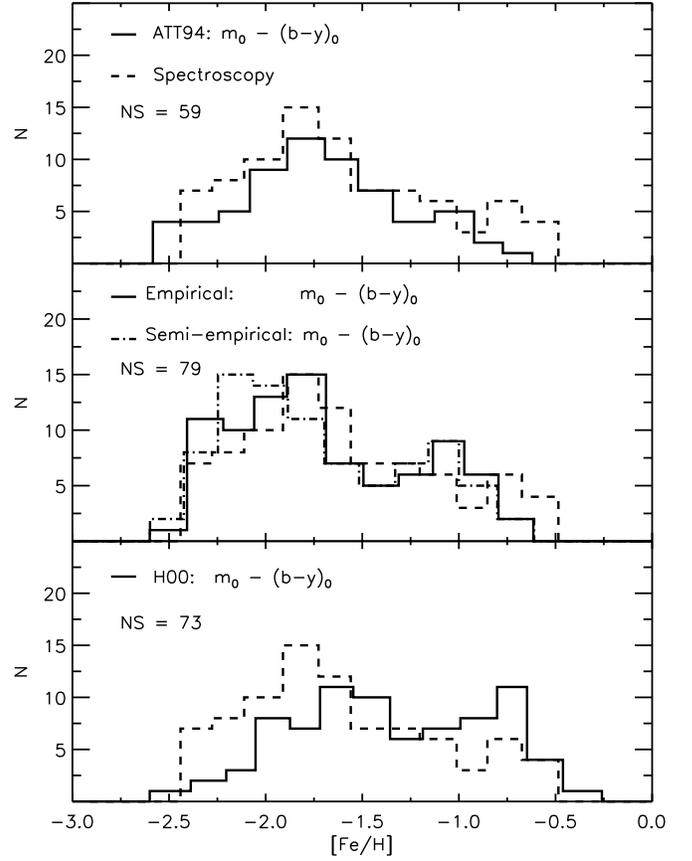}
\caption{ \small{Comparison between spectroscopic and photometric metallicity 
distributions based on the 79 field RG stars from the ATT94 and ATT98 sample 
(dashed line) versus the spectroscopic iron abundance. Top: the solid line 
shows the metallicity distribution based on the ATT94 empirical 
--$m_0, (\bmy)_0$--  MIC relation. Middle: the solid and the dashed-dotted 
lines display the  metallicity distributions based on our empirical and 
semi-empirical --$m_0, (\bmy)_0$-- MIC relations.  Bottom: the solid line 
shows the  metallicity distribution based on the H00 empirical 
--$m_0, (\bmy)_0$-- MIC relation.}
}
\end{center}
\end{figure}
\section{Comparison between photometric and spectroscopic iron abundances}\label{compa}

To estimate the metal content of the selected \omc RG stars we adopted 
the MIC relations provided by CA07 (see their Table~3) and based on the 
$\umy$ and on the $\vmy$ color, together with the new MIC relation 
derived in \S 6. The photometric metallicities were estimated 
using both the empirical and the semi-empirical calibrations.   
\begin{figure}
\label{fig12}
\begin{centering}
\includegraphics[height=0.6\textheight,width=0.5\textwidth]{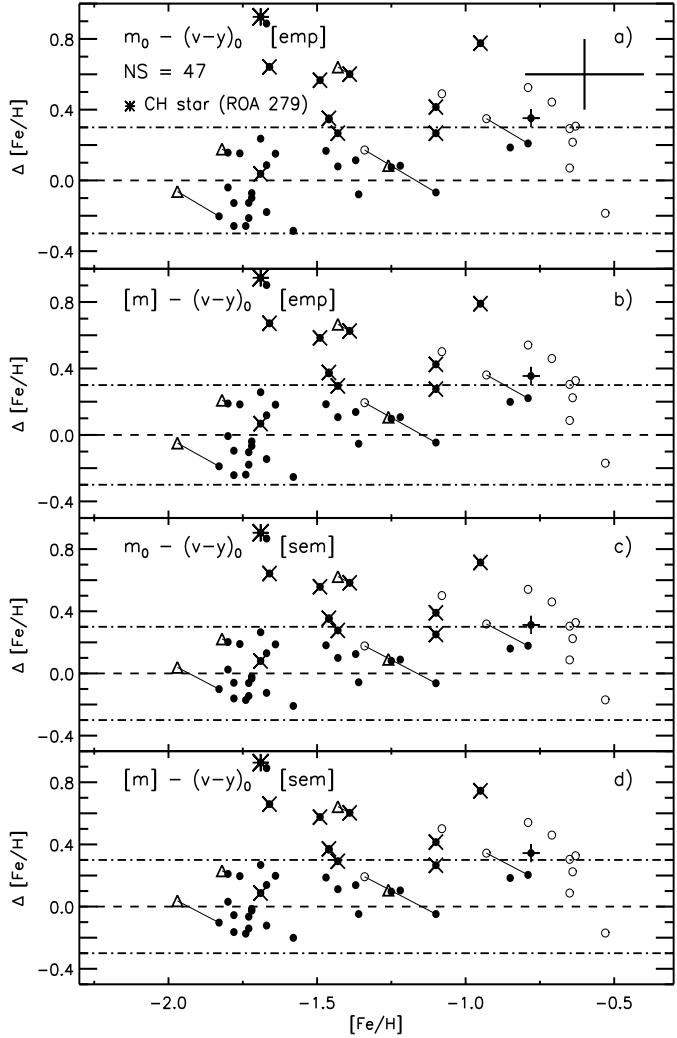}
\caption{
\small{Difference between photometric and spectroscopic metallicities, 
$\Delta \feh \equiv \feh_{phot} - \feh_{spec}$, plotted versus 
$\feh_{spec}$. Spectroscopic iron abundances for 39 ROA stars come 
from the list of ND95 (filled circles), eight from the list of PA02 
(open circles), and four from the list of SM00 (open triangles). 
Solid lines connect RGs with spectroscopic measurements from different 
authors. Panels {\em a)} and {\em b)} display the differences for two 
empirical calibrations ($m_0, (\vmy)_0$; $[m], (\vmy)_0$), while the 
panels {\em c)} and {\em d)} show the differences for the same MIC relations, 
but based on the semi-empirical calibrations. Crosses mark {\rm CN}-strong stars 
according to ND95.  The error bars in the top panel represent photometric 
and spectroscopic errors.}
}
\end{centering}
\end{figure}
The individual metallicities were estimated by assuming for \omc a mean 
reddening value of $E(B-V)$ = 0.11 (CA05) and the extinction 
coefficients for the \strom colors discussed in \S 4.

To verify the accuracy of the metallicity estimates, we cross-correlated our RG
sample with the iron abundances---based on high-resolution spectra---of 40 ROA
stars provided by ND95. We found 39 RG stars in common. Moreover, we matched our 
photometric catalog with the iron abundances---again based on high-resolution 
spectra---of ten metal-rich RGs provided by PA02. Two of these 
stars (ROA~179, 371), are 
also in the ND95 sample. We also adopted the iron abundances 
provided by SM00 for four RGs in common with the ND95 sample 
(ROA~102, 213, 219, 253) based on similar high-resolution 
spectra. We ended up with a sample of 47 RGs with accurate 
iron measurements for which we have either three (47, $vby$) or 
four (28, $uvby$) independent \strom magnitudes.  
The difference between the inferred photometric metallicities and the
spectroscopic measurements are plotted versus spectroscopic 
abundances in Fig.~12.  The four different panels display 
metallicity estimates based on different empirical MIC 
relations. Data plotted in this figure indicate that 
photometric and spectroscopic abundances are in reasonable 
agreement. The mean difference for the four MIC relations is 
$<\Delta \feh > = <\feh_{phot}-\feh_{spec}>$= 0.17$\pm$0.01 dex, 
with a dispersion of the residuals of $\sigma$ = 0.31 dex.  
Once we remove {\rm CN}-strong and chemically peculiar stars 
({\em vide infra}), the scatter around the mean is mainly due 
to photometric and spectroscopic measuring errors. 
The error bars on the top panel of Fig.~12 display the entire error budget. 
Note that throughout the paper with the symbol $\feh_{phot}$ we mean the 
overall stellar metallicity, i.e., the sum of all elements beyond helium
(VandenBerg et al.\ 2000).

Uncertainties in reddening corrections minimally affect the current 
metallicity estimates, and indeed the photometric metallicities based on 
the reddening-free metallicity index, $[m]$, have similar differences 
and dispersions (panels b, d). The same outcome applies to photometric 
metallicities based on empirical and on semi-empirical MIC calibrations.   
However, eight RG stars with iron abundance ranging from $\feh_{spec}=-1.7$ 
to --0.9 
(ROA 53, 100, 139, 144, 162, 253, 279, 480) present larger 
($\Delta \feh >$ 0.3 dex) differences.  Interestingly, six of them 
(marked with crosses in Fig. 12) are {\rm CN}-strong stars, while 
ROA 279 (marked with an asterisk) is a {\rm CH}-star
(see, e.g. ND95). The {\rm CN}-weak star ROA 53 has $\Delta \feh \approx$ 0.8 dex, 
but the photometric metallicity is probably affected by poor photometric 
precision ($\sigma_{v,b,y} \sim$ 0.05). The other three discrepant stars, 
with $\feh_{spec} \ge -1.0$ belong to the PA02 sample. 
However, the PA02 spectroscopic measurements for the two ROA stars in common 
with ND95 are more metal-rich by $\approx$ 0.2 dex. 
This systematic difference has already been discussed by PA02. 
On the other hand, the iron abundances provided by ND95 and SM00 for the 
same RGs agree within the uncertainties.  This suggests that these more 
metal-rich discrepant objects deserve further investigation. 

If we neglect the {\rm CN}-strong stars, the {\rm CH}-star, the ROA53, and the three 
stars from the PA02 sample, the mean difference for the four MIC relations 
is $0.02\pm0.02$ dex, with a dispersion of the residuals of $\sigma$ = 0.16 dex.

   \begin{figure}
\label{fig13}
   \begin{center}
   \includegraphics[height=0.6\textheight,width=0.5\textwidth]{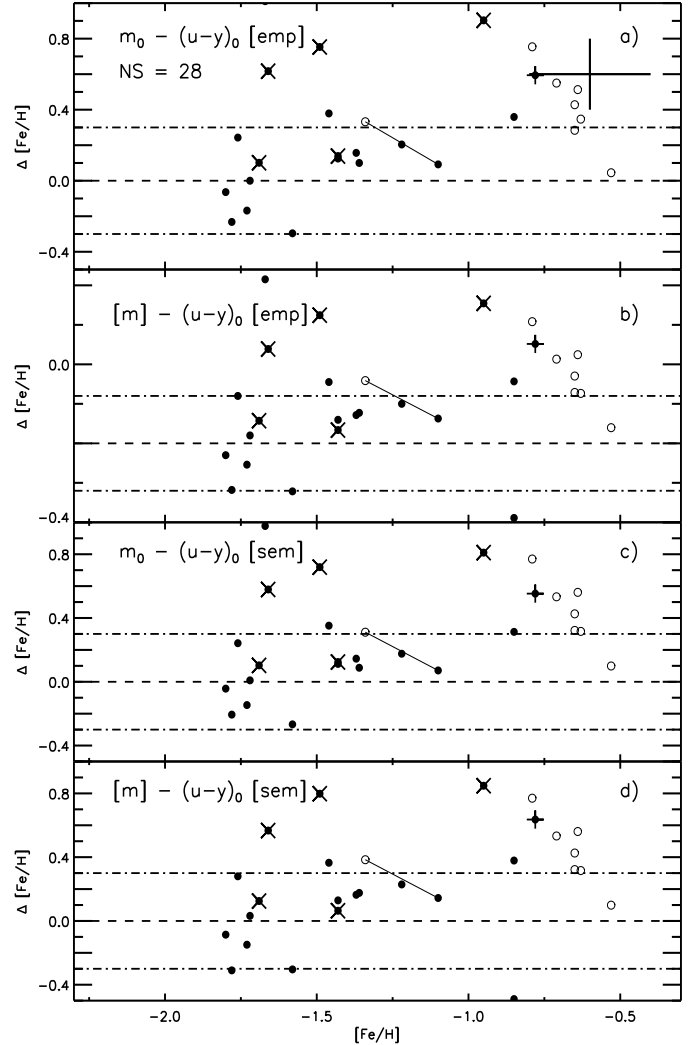}
   \caption{ \small{Same as Fig.~12, but for the 28 stars for which the 
   $u$-band photometry is available. Photometric metallicities are based on 
   four different MIC relations ($m_0, (\umy)_0$; $[m], (\umy)_0$). 
   Crosses mark {\rm CN}-strong stars according to ND95, while the plus marks 
   the candidate {\rm CN}-strong star ROA 248.}
            }
    \end{center}
    \end{figure}

To further constrain the plausibility of our findings, we performed the same 
comparison using 28 stars for which $u$-band photometry is 
available. Fig.~13 shows the difference between photometric and spectroscopic 
abundances using both empirical (panels a,b) and semi-empirical (panels c,d) 
calibrations ($m_0, (\umy)_0$; $[m], (\umy)_0$).
The result is quite similar to the MIC relation based on the $\vmy$ color, 
and indeed the mean difference is $<\Delta \feh>$ = 0.26$\pm$0.02 dex, 
and the dispersion of the residuals is $\sigma \sim$ 0.36 dex. The star 
ROA 248 ($\feh = -0.78$ dex) shows a large discrepancy 
($\Delta \feh\approx 0.6$). However, ROA 248 could be a candidate
{\rm CN}-strong star, as suggested by the large value of the cyanogen index, 
S3839 = 0.51 (ND95), and by its peculiar position in the 
$m_0,(\vmy)_0$ plane and in the $y_0,(\vmy)_0$ CMD (see Fig.~14). 
If we neglect the five {\rm CN}-strong stars, the stars ROA 248, ROA 53, and the 
three stars from PA02, for which the discrepancy in this plane is even 
larger, the mean difference for the four MIC relations is 
$0.07\pm0.03$ dex, with a dispersion of $\sigma$ = 0.16 dex.

We performed the same test using the empirical and the semi-empirical 
$m_0, (\bmy)_0$ MIC relations and we found that the mean difference 
over a sample of 47 objects is 
$<\Delta \feh>$ = 0.22$\pm$0.05 dex, and the dispersion of the residuals 
is $\sigma \sim$ 0.43 dex. Once we remove the objects with peculiar abundances 
we find $<\Delta \feh>$ = 0.01$\pm$0.03 dex, and the dispersion of the 
residuals is $\sigma \sim$ 0.24 dex. 

To avoid subtle uncertainties caused by the limited sample of RG stars 
with accurate spectroscopic measurements, we took advantage of the recent 
large sample of iron abundances, based on high-resolution spectra for 
180 RG stars in \omcp, collected by Johnson et al.\ (2008, hereinafter J08).
This catalog was cross-correlated with the proper motion measurements by 
LE00 and we considered as candidate cluster stars those with a proper motion 
probability of 80\%. We cross-correlated this spectroscopic sample with 
our photometric catalog and we found 118 RGs in common. We estimated 
the difference between photometric ($m_0,\vmy_{emp}$) and spectroscopic 
metallicities and found a mean difference of $0.05\pm0.02$ dex, 
with a dispersion of the residuals of $\sigma$ = 0.36 dex.

To constrain on a quantitative basis the impact of possible {\rm CN} 
enhancements we supplemented previous spectroscopic catalogs with 
the large set of medium-resolution spectra collected by VL07 (for 
more details see \S 9). We found 373 RGs in common with our 
photometric catalog. 
   \begin{figure}[ht!]
\label{fig14}
   \begin{center}
   \includegraphics[height=0.35\textheight,width=0.5\textwidth]{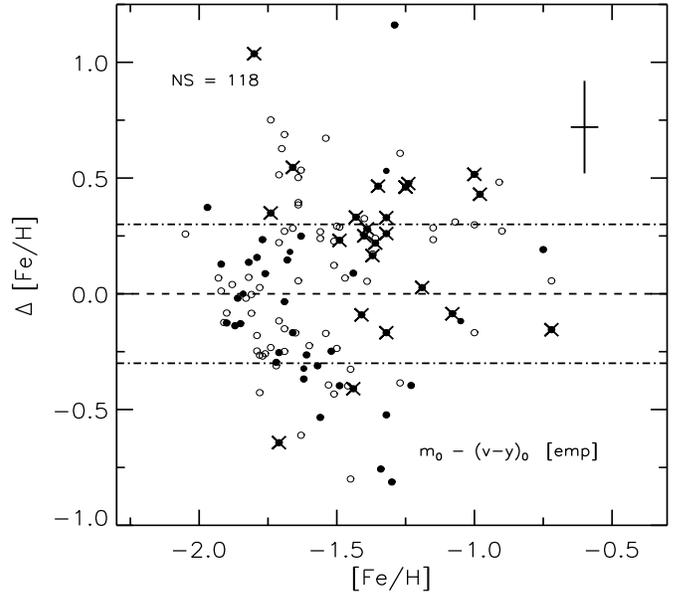}
   \caption{\small{Difference between photometric and spectroscopic 
    metallicities, $\Delta \feh = \feh_{phot}(m_0, \vmy_0)_{emp} - \feh_{spec}$, 
    plotted versus 
    $\feh_{spec}$ for the 118 RG stars with accurate iron abundances by J08. 
    Filled circles mark stars with the measurement of the spectroscopic $S3839$ 
    index (VL07), while those for which this measurement is not available are 
    marked with open circles. Crosses mark {\rm CN}-strong stars according to both  
    our selection (see \S 9) and to ND95. The error bars account for both 
    photometric and spectroscopic errors. The mean difference is 
    $\sim 0.05\pm0.03$ dex, with a dispersion in the residuals of 
    $\sigma = 0.37$ dex. The stars with $\Delta \feh \ge 0.3$ dex are 
    either {\rm CN}-strong or do not have a spectroscopic {\rm CN}-band measurement.
    }
            }
    \end{center}
    \end{figure}

Fig.~14 shows the difference between photometric and spectroscopic iron abundances 
for the 118 RGs in common with J08. Filled circles mark the stars for which 
the measurement of the $S3839$ index (cyanogen band) is available (VL07), while 
the open circles represent stars which lack of this measurement. The crosses mark 
the {\rm CN}-strong stars according to either our selection ($\delta CN > 0.2$, 
see \S 9) or to that of ND95. Data plotted in this figure show that 
{\rm CN}-strong stars are, as expected, concentrated among the more metal-rich 
stars ($\feh \gtrsim -1.5$, Gratton, Sneden \& Carretta 2004).
The mean difference between photometric and spectroscopic abundances, using 
the six different MIC relations, is $\Delta \feh = 0.05\pm0.03$ and 
$\sigma= 0.37$ dex.  Once again we found that the dispersion of the residuals 
is larger in the $m_0, (\bmy)_0$ than in the $m_0, (\vmy)_0$ relation 
( 0.36 vs 0.42, empirical; 0.33 vs 0.40, semi-empirical).    

It is worth mentioning that 25 out of the 118 stars show
discrepancies in iron abundance larger than 0.3 dex. 
The measurement of the  {\rm CN} index is available for thirteen of these stars 
and among them ten are  {\rm CN}-strong. If we move the limit down 
to 0.2 dex, the number of discrepant stars is 46, and the  {\rm CN} index is 
available for 20 of them. The fraction of  {\rm CN}-strong stars becomes 
of the order of 75\%.    
On the other hand, 17 stars appear to be under-abundant in iron by more than 
0.3 dex; among them ten have measurements of the  {\rm CN} index and 
only one is a  {\rm CN}-strong star.  The current findings thus indicate 
that stars with large {\it positive discrepancies\/} in the photometric metallicity 
are highly correlated with the occurrence of strong  {\rm CN} bands. 
The stars with large {\it negative\/} discrepancies require a more detailed 
spectroscopic investigation to constrain their abundance pattern.  
If we neglect the  {\rm CN}-strong stars  we find a mean difference of 
$\Delta \feh = 0.02\pm0.04$ and $\sigma= 0.37$ dex (93 stars).

As a final test of the intrinsic accuracy of the photometric abundances,
we plotted the 118 RGs by J08 (open circles), the 39 ROA stars (open triangles) 
and the 8 RGs by PA02 (open diamonds) onto the $m_0,(\vmy)_0$ plane 
(Fig.~15, top panel). The spectroscopic stars have the expected $m_0$/color 
distribution with the eight metal-rich RGs by PA02 attaining the largest 
$m_0$ values and color the faintest apparent $y_0$ magnitudes at fixed color
(bottom panel).  The plausibility of their metal-rich nature is also 
supported by the fact that six of them are located along the anomalous 
RGB (\omtre). It is noteworthy that  {\rm CN}-strong stars (crosses) possess 
intermediate $m_0$ values and apparent $y_0$ magnitudes. This finding 
should not be affected by selection effects, since the low-resolution 
spectra collected by V07 cover a significant fraction of the RG branch 
in \omc (down to V$\sim$16). The same outcome applies for the {\rm CH}-star
(asterisk), while the candidate  {\rm CN}-strong star ROA~248 (plus) shows the 
largest $m_0$ value and the reddest $\vmy$ color in the sample, thus 
supporting its peculiar nature.  
 
   \begin{figure}
\label{fig15}
   \begin{center}
   \includegraphics[height=0.6\textheight,width=0.5\textwidth]{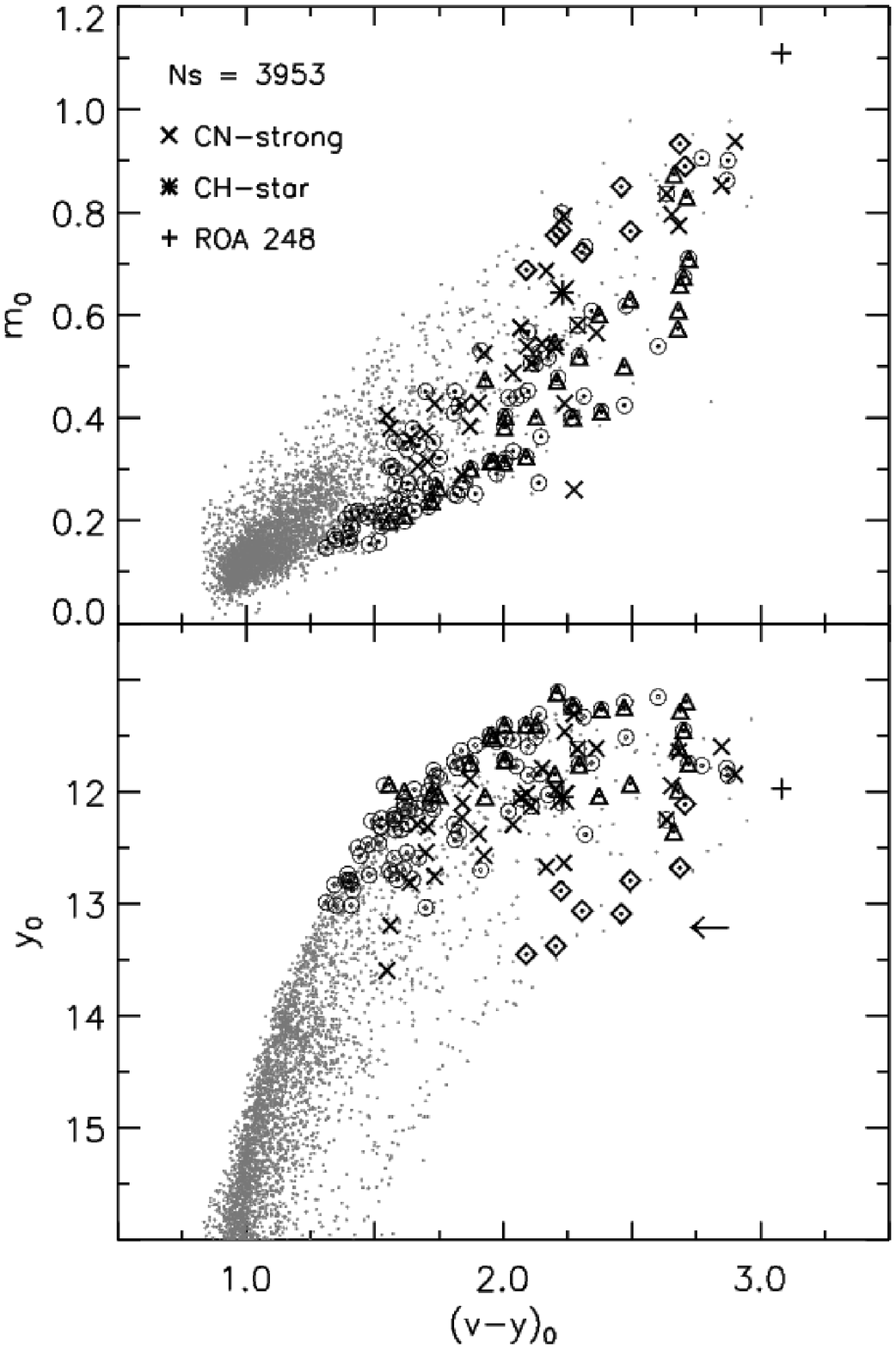}
   \caption{\small{Top: RG stars of \omc in the $m_0, (\vmy)_0$ plane. 
   The open circles mark the 108 RGs from J08, the open triangles mark the 
   39 ROA stars from ND95 and the diamonds the eight RGs from PA02. The 
   {\rm CN}-strong stars are marked with crosses, the {\rm CH}-star (ROA~279) with 
   an asterisk and the candidate {\rm CN}-strong star (ROA~248) with a plus. 
   Bottom: Same as the top, but in the $y_0, (\vmy)_0$ CMD. The arrow 
   indicates the metal-rich RGs located along the \omtre~ branch. 
    }
            }
    \end{center}
    \end{figure}

The above results highlight three relevant findings:  
{\em i)}-- Photometric iron abundances agree quite well with  
spectroscopic measurements, and indeed the mean difference is, 
within the errors, negligible. This supports the hypothesis 
suggested by CA07 (see also \S 6) that the discrepancy between 
photometric and spectroscopic abundances found for field RG stars 
is caused by an intrinsic difference in their abundance patterns.    
{\em ii)}-- The MIC relations based on the $(\umy)$ and the $(\vmy)$ 
color are more affected by {\rm C} and {\rm N} enhancements. Two strong 
 {\rm CN} absorption bands at $\lambda=4142$ and $\lambda=4215$ \AA, 
are close to the effective wavelength of the 
$v$ ($\lambda_{eff}=4110$ \AA) filter, while the two  {\rm CN} bands at 
$\lambda=3590$ and $\lambda=3883$ \AA~ affect the 
$u$ ($\lambda_{eff}=3450$ \AA) filter (see e.\,g.\ Smith 1987; 
Bell \& Gustafsson 1978; ATT94; H00; Grundahl et al.\ 2002; CA07).
Although these MIC relations depend on  {\rm CN} both in the 
$m_1$ index and in the color, they provide more robust photometric 
abundances when compared with the $m_0, (\bmy)_0$ relation, due to   
their stronger temperature sensitivity.  
{\em iii)}-- The dispersion of the residuals between photometric and 
spectroscopic abundances is of the order of 0.3--0.4 dex. However, 
the dispersion decreases by at least a factor of two if all the 
stars appearing to have abundance anomalies (ND95 data) are removed 
from the sample.


\section{Photometric metal abundance distributions}\label{metadis}

In the previous section we have investigated pros and cons of the different
MIC relations to estimate the metal content of RG stars. To constrain the 
metallicity distribution of \omc RGs we selected the stars with photometric 
precision $\sigma_{v,b,y} \leq$ 0.015 and, $\sigma_u \leq$ 0.02 mag,
and $sep>$ 3. Moreover, we excluded all the stars with colors redder/bluer 
than the color range covered by empirical and semi-empirical calibrations (CA07).   
We ended up with a sample of 3,953 RGs; among 
them 2,846 stars also have the $u$-band measurement. Independent estimates 
of the photometric metallicity were obtained from the three different 
MIC relations and the two calibrations\footnote{The selected photometric 
catalogs including colors and metallicity can be retrieved from the following URL:
http://venus.mporzio.astro.it/~marco/spress/data/omegacen/}. 
Fig.~16 shows the six different 
metallicity distributions for the selected RGs. The individual distributions 
were smoothed with a Gaussian kernel having a standard deviation equal to the 
photometric error in the $m_1$ index. 
To provide a robust fit of the metallicity distributions we developed an
interactive program that performs a preliminary Gaussian fit of the main
peaks. On the basis of the residuals between the metallicity distributions
and the analytical fit, the software allow us to insert manually new Gaussian
components. At each step, the code compute a new global solution and the
procedure is iterated until the residuals are vanishing. We fit the
distributions with a sum of seven Gaussians and the dashed lines plotted
in Fig.~16 display the cumulative fits, while the asterisks mark the
position of the different Gaussian peaks. The peaks and the sigma of the
different Gaussian fits are listed in Table~13. 

   \begin{figure*}
\label{fig16}
   \begin{center}
\includegraphics[width=15cm, height=17cm,angle=90]{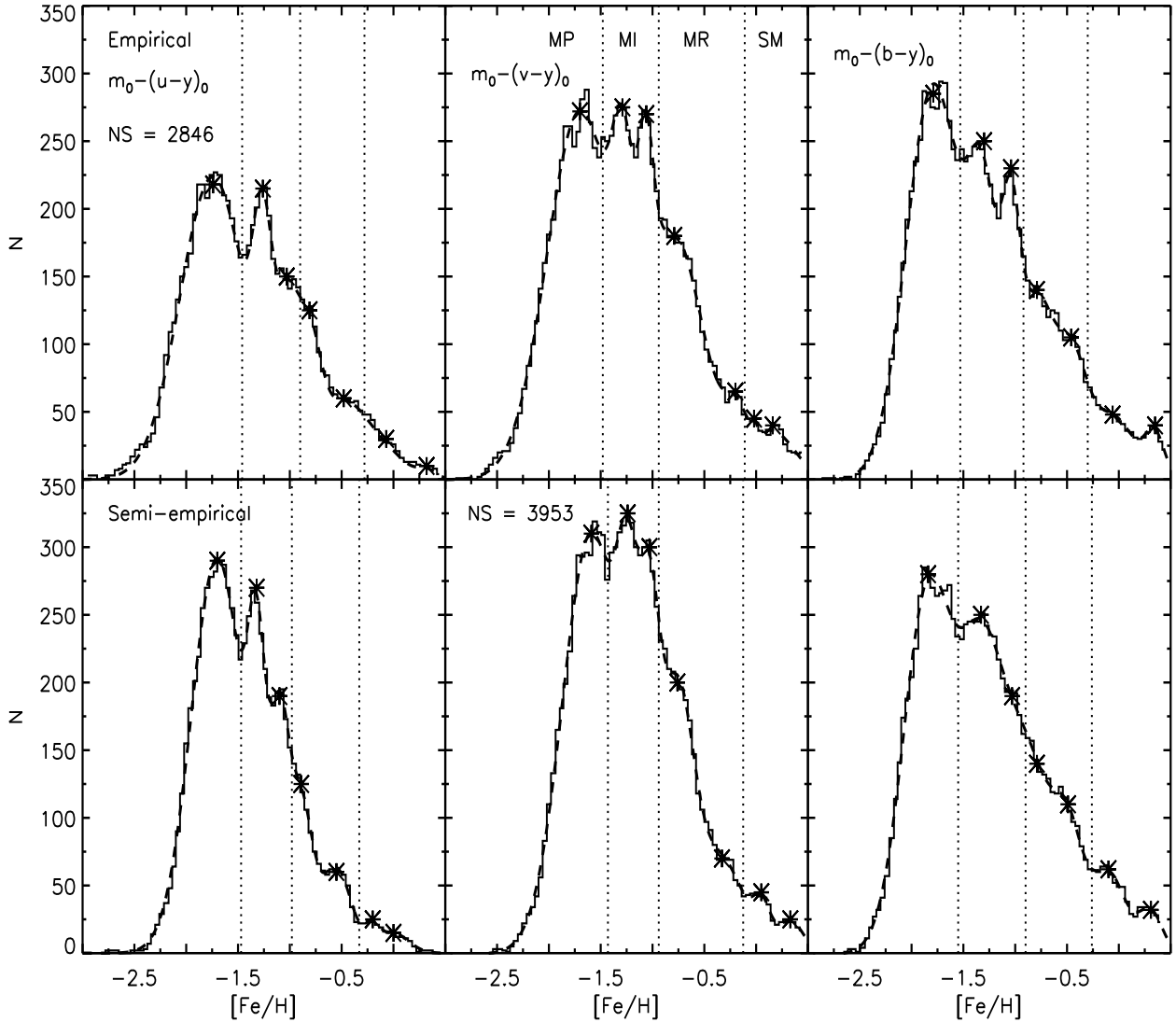}
   \caption{\small{Distribution of photometric iron abundances based on 
    empirical (top) and semi-empirical (bottom) calibrations of different 
    MIC relations. The sample size of the $m_0, (\umy)_0$ (left), of the 
    $m_0, (\vmy)_0$ (middle) and of the $m_0, (\bmy)_0$ (right) relations 
    are also labeled. The metallicity 
    distributions were smoothed with a Gaussian kernel with a $\sigma$ equal 
    to the individual intrinsic error on the $m_1$ index. The dashed line 
    shows the fit of the metallicity distribution computed as the sum of 
    seven Gaussians. The asterisks mark the peaks of the seven Gaussians. 
    The vertical dotted lines display the different metallicity regimes: 
    metal-poor (MP), metal-intermediate (MI), metal-rich (MR) and  
    solar-metallicity (SM).}
            }
    \end{center}
    \end{figure*}

The shape of the metallicity distributions is, as expected, asymmetric with 
a sharp cut-off in the metal-poor tail ($\feh_{phot} <$ --2) and a metal-rich 
tail approaching Solar iron abundances.  The different MIC relations and the 
different calibrations show, within the errors, very similar metallicity 
distributions. The same conclusion applies to the peaks in the metallicity 
distribution (see Table~13). The peaks according to the six 
calibrations are located at $\feh_{phot}=$ $-1.73\pm0.08$, $-1.29\pm0.03$, 
$-1.05\pm0.02$, $-0.80\pm0.04$, $-0.42\pm0.12$ and $-0.07\pm0.08$ dex, 
where the uncertainties are the standard errors of the mean. A handful of objects 
seems to show super-Solar iron abundance 
-- $\feh_{phot}\approx$ $0.24\pm0.13$ --, 
but this sample vanishes in the $m_0,\ (\umy)_0$ distribution and 
might be caused by  {\rm CN}-strong stars of the metal-rich peak at
$\feh_{phot}\approx $ -0.07 dex. 
These findings are minimally affected by uncertainties in reddening 
corrections, since the metallicity distributions based on the reddening-free
$[m]$ index are very similar (Calamida 2007).

The four most significant peaks ($-1.73 \le \feh \le -0.80$) in the photometric 
iron distributions agree quite well with the spectroscopic peaks measured 
by J08 (see their Fig.~10). Iron abundances based on high-resolution 
spectra also suggest the presence in \omc of more metal-rich stars
($\feh =-0.60\pm0.15$) stars (ND95; PA02; Pancino 2004).
The outcome is the same if we compare photometric abundances with
metallicity distributions based on low and medium resolution spectra. 
These investigations are typically based on measurements of the 
calcium triplet (N96, 517 RGBs; SK96, 343 RGBs;
S05b, 152 SGBs; Stanford et al.\ 2006a, 442 MSs)
that are transformed into iron abundances via empirical relations,
or on direct iron line measurements (H04, 
$\sim$ 400 MSs and SGBs; van Loon et al.\ 2007, hereinafter VL07,
$\sim$ 1500 HBs and RGBs; Villanova et al.\ 2007, 80 SGBs).
In particular, the metal-rich tail in the metallicity distributions 
based on larger cluster samples approaches $\feh \approx -0.5\pm0.2$ dex.
The metallicity distributions based on photometric indices 
(HR00, $m_1$-index, $\sim1500$ RGs; Rey et al.\ 2000,
$hk$-index 131 RR Lyrae; Hughes et al.\ 2004, $m_1$-index, $\sim 2500$
MSs, SGBs, RGBs) also provide similar results.

   \begin{figure}
\label{fig17}
   \begin{center}
\includegraphics[height=0.45\textheight,width=0.5\textwidth]{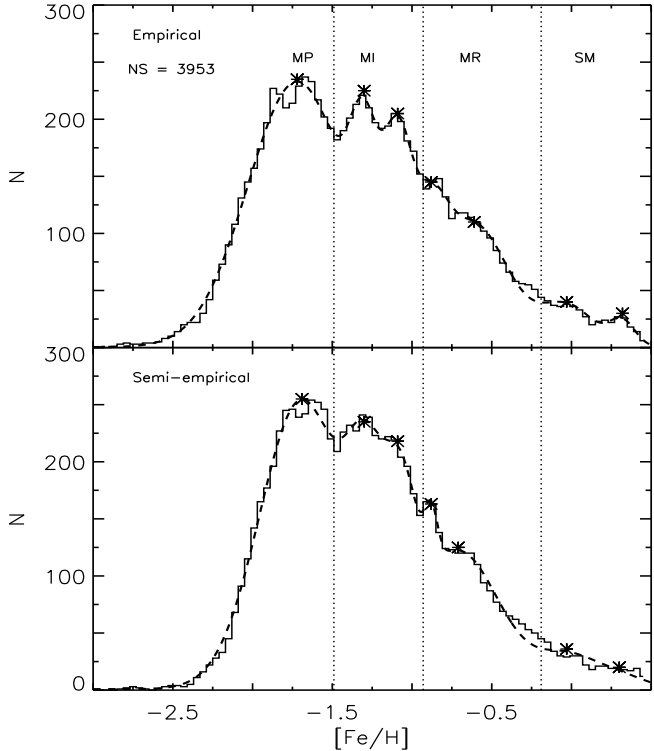}
   \caption{\small{Average distribution of photometric iron abundances 
    based on the three empirical (top) and on the three semi-empirical 
    (bottom) calibrations of the adopted MIC relations. The asterisks 
    mark the peaks of the seven Gaussian adopted in the cumulative fit 
    (dashed line). The vertical dotted lines display the mean of the 
    different metallicity regimes identified in Fig.~16.}
            }
    \end{center}
    \end{figure}
The large number of RGs allows us to identify 
four different metallicity regimes: 
metal-poor (MP) with $\feh \lesssim -1.5$, 
metal-intermediate (MI) with $-1.5 \lesssim \feh \lesssim -1.0$, 
metal-rich (MR) with $-1.0 \lesssim \feh \lesssim -0.1$ and 
Solar metallicity with $\feh \approx 0$ dex (see vertical dotted 
lines in Fig.~16). The limits of these metallicity regimes 
are arbitrary, though identified with the occurrence of either a 
local minimum or a shoulder in the metallicity distribution.  
  
Following the anonymous referee suggestion we estimated the average metallicity 
distributions of both empirical and semi-empirical MIC relations (see Fig.~17). 
We performed on the average distributions the same Gaussian fits adopted for 
the individual distributions. Data plotted in Fig.~17 (see also Table~13) 
show that both the peaks and the metallicity regimes are, within the errors, 
quite similar. The difference between the cumulative fits and the metallicity 
distributions suggests the possible occurrence of a secondary peak located 
at the edge between the MR and SM regime ($[Fe/H]\sim -0.25$).  

The main findings concerning the metallicity distributions in \omc
according to the photometric abundance estimates are the following:

{\em i)} The metallicity distribution shows a very sharp cutoff
in the metal-poor regime, and the fraction of stars more metal-poor than
$\feh \approx -2.0$ dex is vanishing.

{\em ii)} The main iron peak is located at $\feh \approx -1.73\pm0.08$ dex,
and the weighted mean fraction of stars in this metal-poor regime ($\feh \le -1.49$) 
is $\sim39\pm1$\%. However, four out of the six metallicity distributions
show either a double peak or a shoulder ($m_0,(\umy)_0$; $m_0,(\vmy)_0$, 
see Fig.~16). The distance between the two peaks is on average of the order 
0.2 dex. The large number of stars per bin and the minimal impact of 
 {\rm CN}-strong stars in this metallicity range suggest that the split is real.  

{\em iii)} The metal-intermediate regime ($-1.49 < \feh \le -0.93$) includes two secondary 
peaks ($\feh \approx -1.29\pm0.03$ [MI1], $-1.05\pm0.03$ [MI2] dex) and a 
weighted mean fraction of stars of  $\sim 32\pm 1$\%. 

{\em iv)} The metal-rich regime ($-0.93 < \feh \le -0.19$) also includes two secondary 
peaks ($\feh \approx -0.80\pm0.04$ [MR1] $-0.42\pm0.13$ [MR2] dex) and has a 
weighted mean fraction of stars of $\sim\; 19\pm4$\%. In the spectroscopic 
metallicity distributions available in the literature, the last two peaks 
become more evident as soon as the sample size becomes of the order of 
several hundred stars. 

{\em v)} In the solar metallicity regime ($\feh \approx 0$) two secondary peaks 
are also present ($\feh \approx -0.07\pm0.08$ [SM1], $0.24\pm0.13$ [SM2] dex), but 
only with a small fraction of stars ($\sim\; 8\pm5$\%). This tail should be treated 
with caution, since we still lack firm {\it spectroscopic\/} evidence for the 
presence of RGs with solar iron abundances in \omc. Moreover, we cannot exclude 
the possibility that these objects either are affected by strong {\rm C} and/or 
{\rm N} enhancements, or belong to the Galactic field population.

{\em vi)} The difference in iron content among the individual peaks 
is roughly constant and equal to a factor of 3--4 (--1.7, --1.2, --0.6, and 0
on the logarithmic abundance scale).

{\em vii)} The fraction of stars belonging to the different metallicity 
regimes steadily decreases when moving from the metal-poor to the 
metal-rich domain. The relative fractions based on the different photometric diagnostics 
are in fair agreement (see Table~14). There are three exceptions.
a) The $m_0, (\umy)_0$ indicates a larger fraction of MP stars when compared 
with the other MIC relations ($48\pm2$\% vs $38\pm1$\%). The difference might 
be due to the stronger sensitivity of this color in the faint RG limit 
(see left panel of Fig.~18 and Fig.~11 in CA07). To validate this working 
hypothesis we estimated the same fractions using only bright RGs ($y_0 \le 15$) 
and we found that the new values, within the uncertainties, agree 
quite well ($52\pm2$\% vs $49\pm2$\%). This evidence suggests that the 
current fraction of MP stars should be considered as a lower limit. 
b) The $m_0, (\vmy)_0$ indicates a larger fraction of MR stars when compared 
with the other MIC relations ($24\pm1$\% vs $17\pm1$\%). The difference is 
almost certainly due to the stronger sensitivity of this color to  {\rm CN}-strong 
stars. This hypothesis is also supported by the slow decrease that the metallicity
distribution shows in this abundance interval (see the middle panels in Fig.~16).  
c)  The $m_0,(\bmy)_0$ diagram indicates a significantly larger fraction of SM stars when 
compared with the other MIC relations ($14\pm1$ vs $6\pm1$\%). The difference is
due to the lesser sensitivity of this color in the metal-rich regime (see 
right panel in Fig.~18).  

   \begin{figure*}[ht!]
\label{fig18}
   \begin{center}
   \includegraphics[height=0.5\textheight,width=0.85\textwidth]{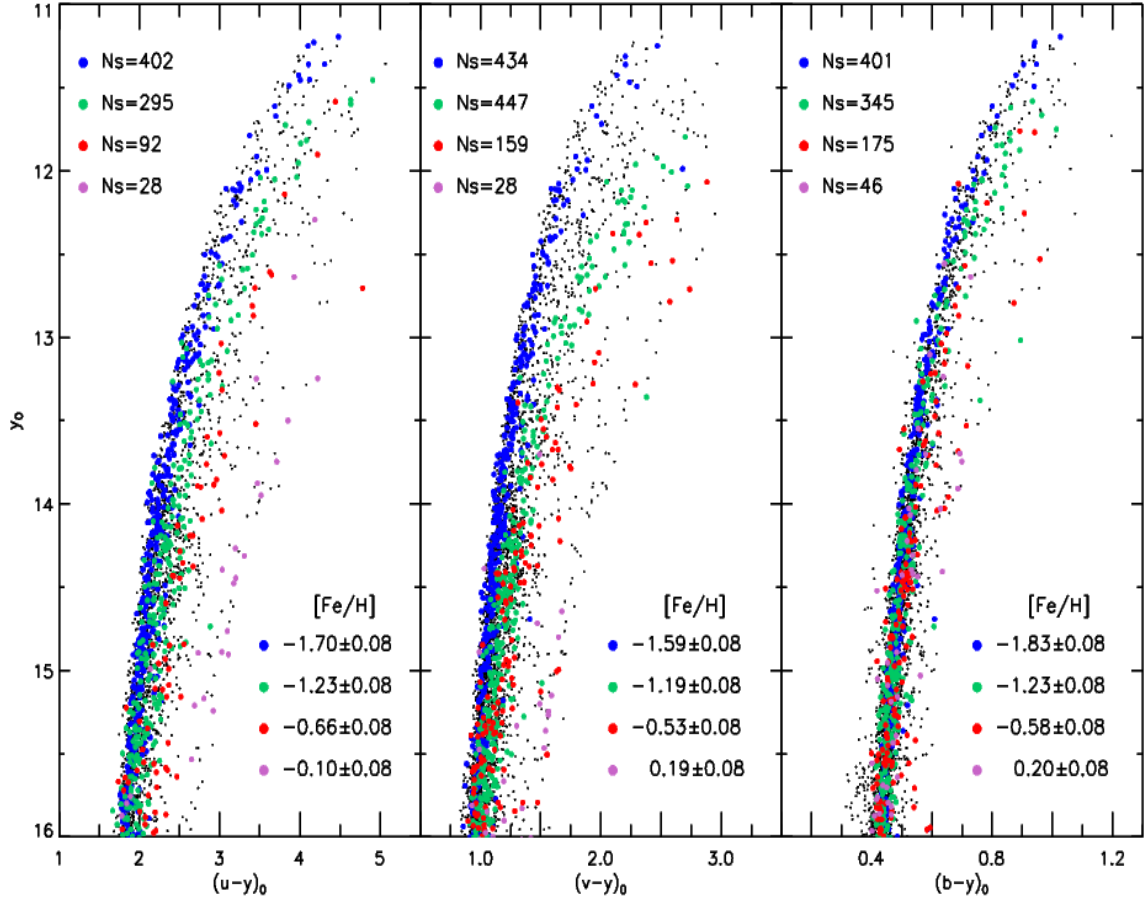}
   \caption{\small{Unreddened \strom CMDs for RGs characterized by different mean 
   iron abundances. The blue dots mark metal-poor (MP) RGs, the green dots 
   metal-intermediate (MI) RGs, the red dots metal-rich (MR) RGs and the purple 
   dots the solar metallicity (SM) RGs. The individual sample sizes and the 
   metallicity intervals are labeled.
    }
            }
    \end{center}
    \end{figure*}

Current findings agree well with the metallicity distribution function obtained
by S05a using broad band photometry of RG stars (1364). In particular, they 
found a well defined MP peak ($Fe/H]\approx$-1.6 dex), three MI peaks 
($Fe/H]\approx$-1.4,-1.1,-0.9 dex) and a sharp peak in the MR tail 
($Fe/H]\approx$-0.6 dex, see their Fig.~8). The same outcome applies 
to the relative fraction of RG stars in the quoted metallicity regimes 
(see their Table~2). 

To test the plausibility of the selected metallicity regimes, Fig.~18 shows 
the distribution of four different stellar groups with different mean metal 
abundances in three different CMDs. The MP sample are RGs located around the
main peak, i.e.,  $\feh \approx -1.70\pm0.08$, while the other samples
are located near other three main peaks: MI for 
$\feh \approx -1.23\pm0.08$,  MR for $\feh \approx -0.66\pm0.08$ and
SM for $\feh \approx -0.1\pm0.08$ dex. Data plotted in the left panel 
show a well defined $\umy$-color ranking over the entire magnitude range 
when moving from the metal-poor (blue dots) to the metal-intermediate 
(green dots), metal-rich (red dots) and solar metallicity 
(purple dots) RGs. The result is the same for the $\vmy$--color (middle 
panel), but the solar metallicity stars located along the \omtre~ branch 
show a larger color scatter and the four different groups merge in 
the faint magnitude range ($y_0 < 15$). The degeneracy becomes more 
evident when moving to the $\bmy$--color (right panel).

The anonymous referee raised the problem that the {\rm CN} molecular bands affect 
the Stroemgren {\rm u} and {\rm v} bands. Therefore, the Stroemegren MIC relations cannot 
be adopted to estimate {\rm [Fe/H]} values in complex systems unless one has a knowledge 
of the strength of those {\rm CN} bands. However, some circumstantial evidence move 
against this working hypothesis.\\  
{\em i)}-- In this investigation we are not attempting to estimate an iron content  
that is identical to the iron abundance measured by spectroscopist using high-resolution 
spectra. We have clearly stated that with {\rm [Fe/H]}, we mean a global metallicity, 
i.e., the sum of all elements beyond helium without explicit assumptions concerning 
the relative distribution among those elements. Plain physical 
arguments indicate that the sensitivity of the Stroemgren indices to metallicity comes 
from changes to the stellar structure ---such as envelope opacity, and consequently 
stellar radius as a function of luminosity---and not merely from atmospheric spectral 
features. For this reason, a photometric index representing "all metals" is a crude 
zero-th order estimate of the chemical content of a star. These metallicity estimates 
are adopted  to rank RG stars according to their heavy element content and to detect 
sample of stars characterized by different relative "metal" abundances.\\  
{\em ii)}--The occurrence of strong molecular bands increases the dispersion in the 
mapping of heavy element abundances. However, even though the Stroemgren indices do 
not provide an exact equivalence with the iron abundance they still give a meaningful 
correlation with the star global metallicity. The precence of this correlation is 
supported by these evidence: {\em a)}-- metallicity distributions based on MIC relations, 
with different sensitivities to the {\rm CN} molecular bands, show similar peaks and 
valleys (see Fig.~16); {\em b)}-- the position of peaks and valleys in the mean 
metallicity distributions (see Fig.~17) are, within the errors, very similar to the 
individual ones; {\em c)}-- the RG stars show a well defined drift in color in the CMD, 
when moving from metal-poor to metal-rich stars (sse Fig.~18); {\em d)}-- the main peaks 
of current metallicity distributions agree quite well with spectroscopic measurements. 
These findings indicate that metallicity estimates based on Stroemgren MIC relations are 
a robust diagnostic to identify stellar populations characterized by different global 
metallicities.\\ 
{\em iii)}-- Accurate spectroscopic measurements indicate variations in the heavy 
element abundance on a star-by-star basis in all GCs. Together with changes in the 
relative abundances of {\rm CNO} elements, well defined anticorrelations have been 
found between {\rm O} and {\rm Na} and between {\rm Mg} and {\rm Al}. Field RG stars 
do not show the above chemical anomalies (Gratton, Sneden, \& Carretta 2004). 
To provide homogeneous metallicity estimates of cluster RGs we provided new calibrations 
of the Stroemgren metallicity indices using cluster RG stars. 
These calibrations appear to be more appropriate in dealing with cluster RGs 
than similar calibrations available in the literature. However, a detailed
comparison between photometric and spectroscopic ($CNO$, $\alpha$-elements)
metallicities is required before reaching a firm conclusion.

\section{Abundance anomalies among RG stars}\label{compa}

\subsection{{\rm CN}-strong stars}\label{compa}

To further constrain the evolutionary and chemical properties of \omc RGs,
we cross-correlated our photometric catalog with the large set (1,519 stars) 
of low-resolution spectra recently collected by VL07. These spectra were  
cross-correlated with the proper motion measurements by LE00 and candidate 
cluster stars were selected according to a membership probability larger 
than 90\%. We found 373 RG stars in common. Fig.~19 shows 
the $S3839$ ({\rm CN}) index for these 
stars\footnote{The $S3839$ index and the {\rm CN} index share the same
definition and measure the strength of the $\lambda$3883 {\rm CN}-band.
They should not be confused with the  {\rm CN}-band located at
$\lambda$4215 (see Smith 1987).} measured by VL07 following the
definition of Norris et al.\ (1981).
To select the  {\rm CN}-strong stars, we adopted 
the $\delta${\rm CN} parameter---the  {\rm CN} excess according to the  
definition by Smith (1987). We defined as  {\rm CN}-strong stars the objects that 
in the  {\rm CN},$(y-K)_0$ plane attain $S3839$ values larger ($\delta CN >$ 0.2) 
than the reference baseline  {\rm CN} = $0.158\times (V-K)_0 - 0.332$\footnote{We are 
assuming, as usual, that the $y$ and the $V$-band magnitudes are effectively 
the same (Crawford \& Barnes 1970).}.
The baseline is based on metal-poor synthetic spectra computed by Cohen 
\& Bell (1986). 
We ended up with a sample of 181 stars shown as crosses 
in Fig.~21. The above selection is arbitrary and driven by the evidence that 
the distribution of  {\rm CN} stars has a local minimum along the dotted line.  
A similar effect is typically detected in GCs showing a bimodal  {\rm CN} distribution  
(Norris et al. 1981; Smith 1987; Kayser et al.\ 2008).

   \begin{figure}[ht!]
\label{fig19}
   \begin{center}
   \includegraphics[height=0.4\textheight,width=0.5\textwidth]{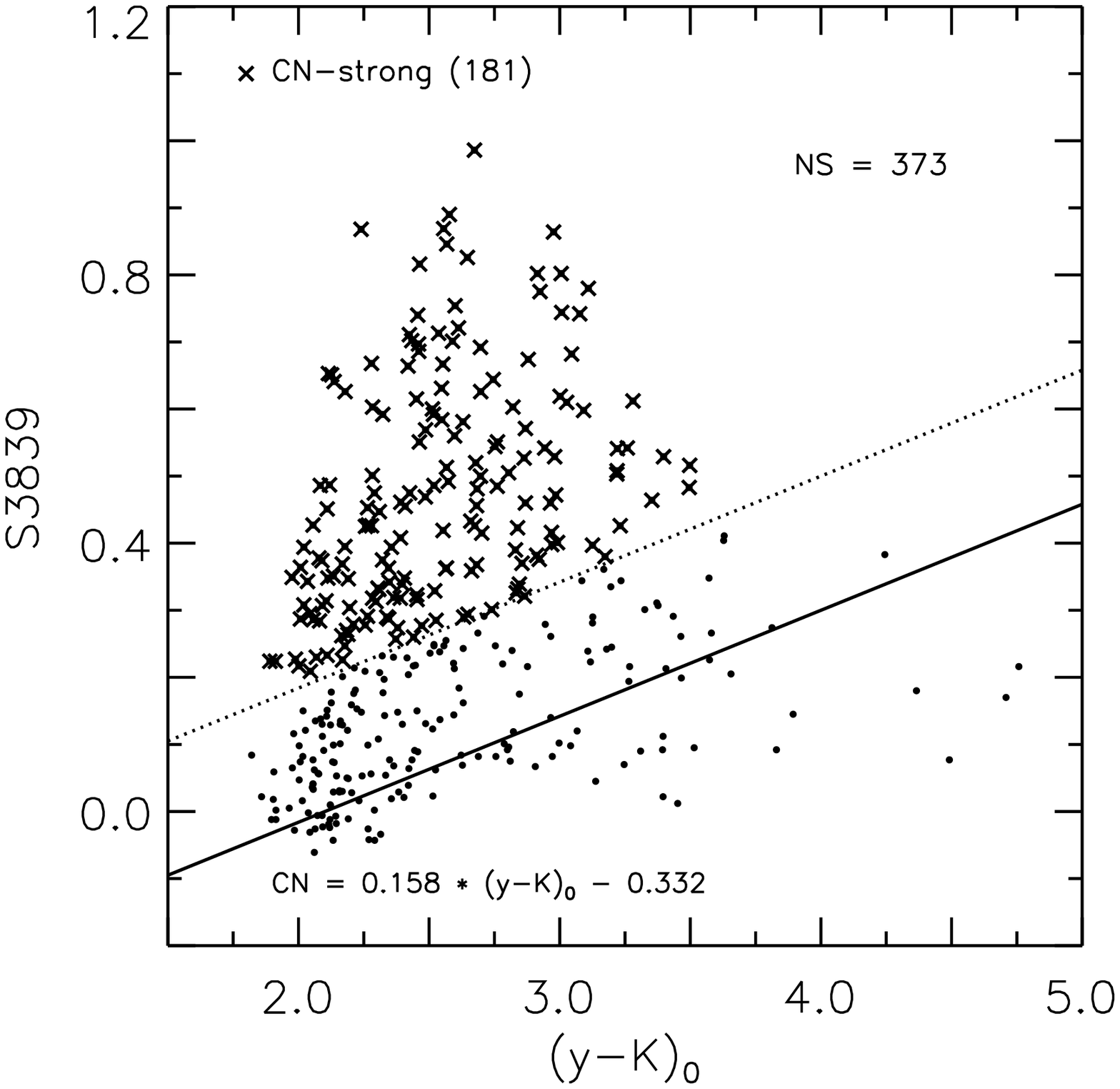}
   \caption{\small{Cyanogen $S3839$ index estimated by VL07 as a function of 
   the unreddened $(y-K)_0$ color. The solid line shows the reference line 
   defined by Smith (1987), while the dotted line the cut we adopted to pin 
   point the candidate {\rm CN}-strong RG stars (crosses, $\delta CN >$ 0.2).}
            }
    \end{center}
    \end{figure}


To constrain the properties of  {\rm CN}-strong stars, Fig.~20 shows their distribution 
in the $S3839$ vs \feh plane. The iron abundances are based on both 
photometric ($\feh_{m_0, (\vmy)_0}$, top; $\feh_{m_0, (\umy)_0}$, bottom) and 
spectroscopic (J08, ND95) estimates. These data support the evidence suggested 
by VL07 that only a few  {\rm CN}-strong stars belong to the metal-poor tail 
($\feh < -1.75$), while the peak of the distribution is located around 
$\feh\approx -1$ and extends to Solar metallicity values. It is worth noting 
that  {\rm CN}-weak and  {\rm CN}-strong stars appear well separated, and indeed 
only a few objects are located at $S3839\approx 0.18$. This supports the criterion 
we adopted to distinguish  {\rm CN}-strong stars. Data plotted in Fig.~20 also show 
that  {\rm CN}-weak stars range from $\feh \approx -2$ to  $\feh \approx -0.75$
(see also  KA06), while more metal-rich RGs seem to be 
 {\rm CN}-strong stars.

   \begin{figure}[ht!]
\label{fig20}
   \begin{center}
   \includegraphics[height=0.6\textheight,width=0.5\textwidth]{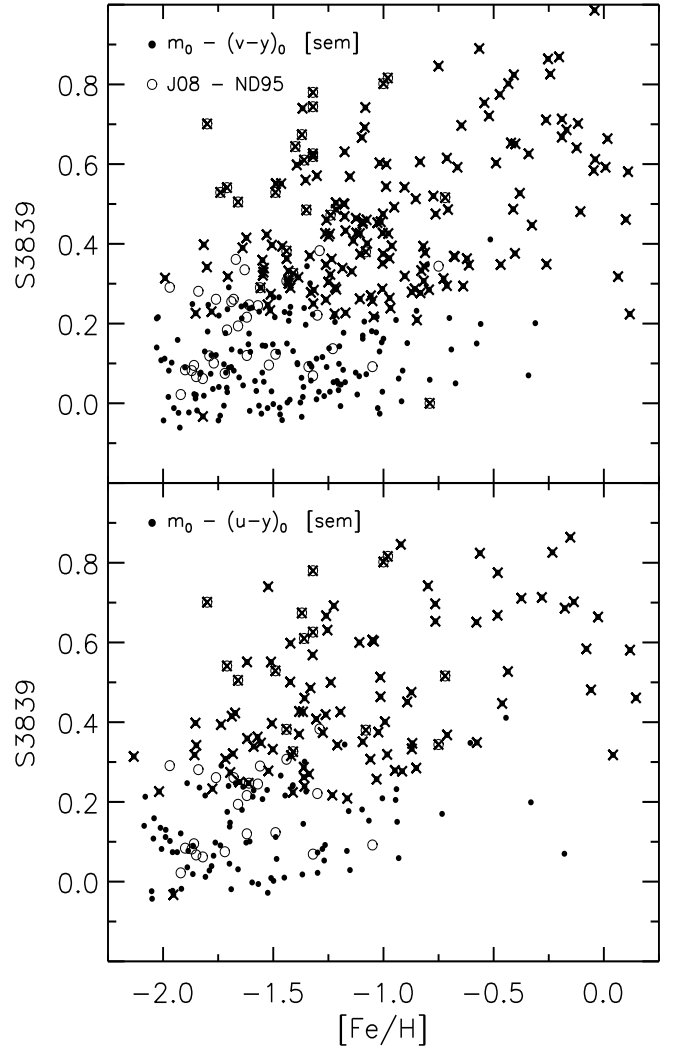}
   \caption{\small{Cyanogen $S3839$ index (filled circles) estimated by VL07 as 
   a function of two different photometric metallicities based on the 
   $m_0, (\vmy)_0$ (top) and on the $m_0, (\umy)_0$ (bottom) semi-empirical MIC relations. 
   Open circles 
   mark the stars with spectroscopic iron abundances by J08 (54) and by ND95 (9). 
   The {\rm CN}-strong stars are marked with crosses. See text for more details.}
            }
    \end{center}
    \end{figure}


As a whole, we find that the fraction of  {\rm CN}-strong stars among the 
objects with measured cyanogen indices is of the order of 50\%.     
However, this fraction is affected by the adopted selection criteria 
and by possible selection effects in the VL07 sample. To provide 
an independent estimate of the relative frequency of candidate 
 {\rm CN}-strong stars, we computed 
the difference between the metallicity estimates based on the 
$m_0, (\vmy)_0$ (more sensitive to the {\rm CN} strength) and the 
$m_0, (\umy)_0$ semi-empirical MIC relations. 
We found that $19\pm1$\% (540 objects)  
of RG stars display a discrepancy larger than 0.2 dex. 
Unfortunately, the measurement of the $S3839$ index (VL07) 
is available for only 15 of these candidate  {\rm CN}-strong stars, but
among them 14 were selected as candidate  {\rm CN}-strong 
stars according to their positions in the $S3839$ vs $(y-K)_0$ plane.   
This means that the above fraction of candidate  {\rm CN}-strong stars should 
be considered as a conservative estimate, due to the ubiquitous effects 
of the  {\rm CN} molecular bands. The current finding, within the errors, agrees 
quite well with the fraction of SGB stars that according to 
Stanford et al.\ (2007) display an enhancement in either {\rm C} ($\sim$ 17\%) or 
N ($\sim$ 16\%; see their Table~4).

   \begin{figure}
\label{fig21}
   \begin{center}
   \includegraphics[height=0.4\textheight,width=0.5\textwidth]{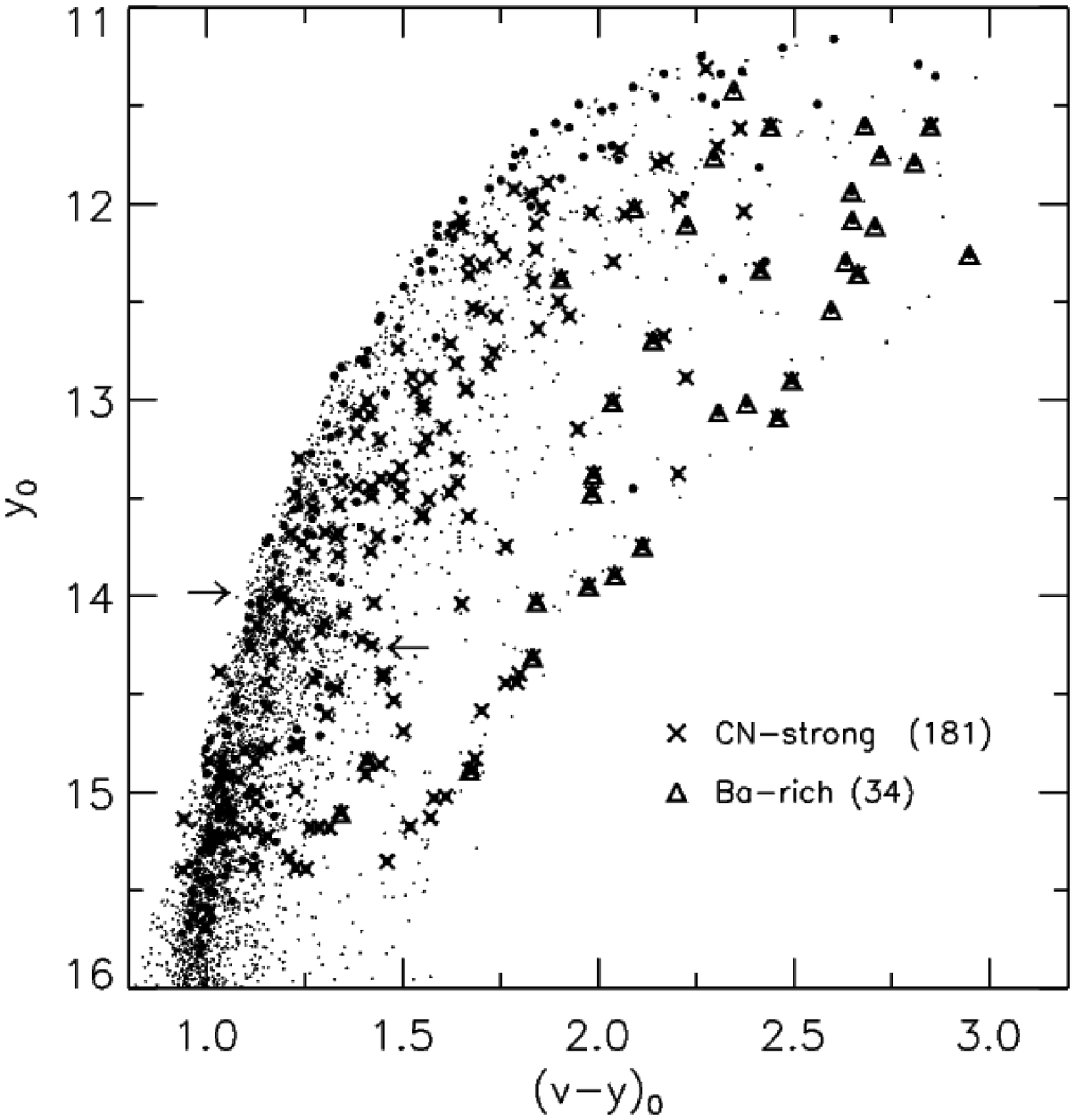}
   \caption{\small{Unreddened $y, \vmy$ CMD of bright RGs. Objects
    in common with VL07 are plotted as filled circles. The crosses mark 
    candidate {\rm CN}-strong stars ($\delta CN > 0.2$, see text for more details), 
    while the triangles the $Ba$-rich stars according to the selection by VL07.
    The two arrows display the region covered by RG bump stars.}  
            }
    \end{center}
    \end{figure}

In passing we note that  {\rm CN}-strong stars are distributed over the entire 
magnitude range covered by our RG sample (Fig.~20). In particular, 
they are present above, below and through the RG bump region (horizontal arrows
in Fig.~21). This further supports the evidence that these objects are 
related to MS and SGB stars apparently having enhancements in either 
{\rm C} or {\rm N}.  Finally, 
Fig.~21 shows the barium-rich stars according to the selection by 
VL07. The clear correlation of these objects with the metal-rich 
RG stars located along the \omtre~ branch strongly supports the 
recent findings by McDonald et al.\ (2008).

\subsection{$\alpha$-element abundances}\label{compa}

We took advantage of the accurate $\alpha$-element 
(O, Mg, Si, Ca, Ti) measurements of RG stars in \omc 
provided by ND95 (filled circles) and by PA02 (open circles) to 
investigate whether the photometric metallicities correlate with 
the abundance of these elements. 
We performed several tests and found that metallicities based 
on the $m_0,(\vmy)_0$ and $m_0, (\umy)_0$ relations show 
very well defined correlations with the abundance of {\rm [Ca+Si/H]}.   
Data plotted in Fig.~22 show that for metal abundances ranging 
from $\feh \sim -1.6$ up to Solar metallicity they display a tight 
correlation. This correlation does not apply to more metal-poor 
objects, and indeed for {\rm [Ca+Si/H]}$\approx$--1.5 and 
$\feh \lesssim -1.7$ dex they form a plateau.  

\begin{figure}
\label{fig22}
   \begin{center}
   \includegraphics[height=0.5\textheight,width=0.5\textwidth]{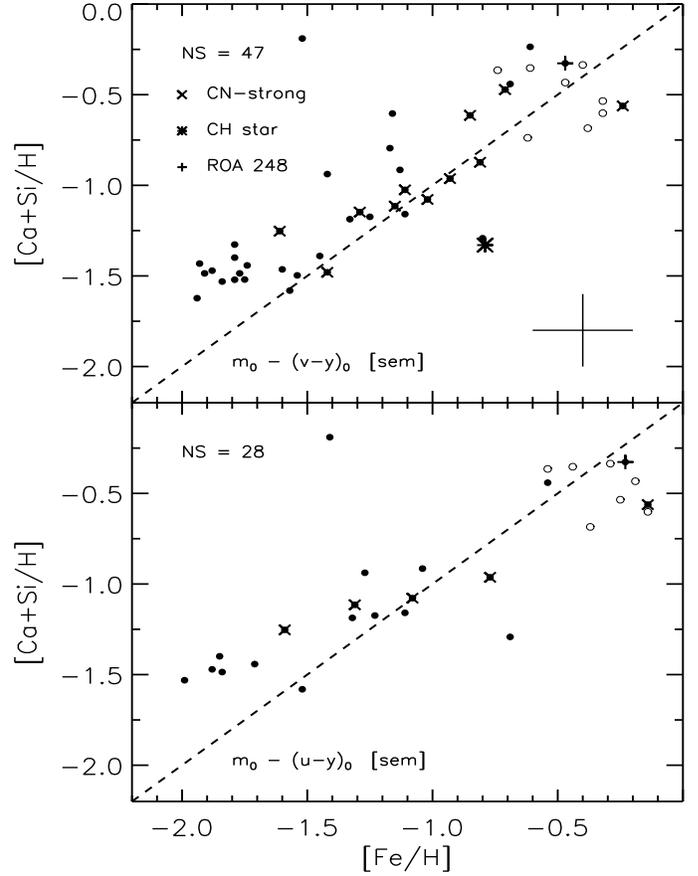}
   \caption{\small{{\rm [Ca+Si/H]} spectroscopic abundances for 
   RG stars in common with ND95 (filled circles) and PA02 (open circles) 
   as a function of photometric metallicities based on the $m_0, (\vmy)_0$ (top), 
   and $m_0, (\umy)_0$ (bottom) semi-empirical MIC relations. The dashed lines 
   display the bisector lines. Crosses mark {\rm CN}-strong stars, the asterisk 
   the {\rm CH}-star and the plus the candidate {\rm CN}-strong star ROA 248. 
   The error bars account for both photometric and spectroscopic errors.}
            }
    \end{center}
    \end{figure}

A good fraction of candidate  {\rm CN}-strong stars follows the same correlation,
while the {\rm CH}-star (asterisk) appears heavily depleted in {\rm Ca+Si} abundance.   
A more detailed analysis concerning the possible occurrence of stars showing
either a {\rm C,N} and/or an $\alpha$-element enhancement or a large {\rm CH} line
blocking would require a larger sample of stars with accurate individual elemental
abundances. 
Note that the correlation between the $\alpha$-elements and photometric metallicities 
might open new opportunities to discriminate between field and cluster RGs.    

\section{Summary and conclusions}\label{compa}

We have presented new and accurate multiband \strom ($u,v,b,y$) and NIR ($J,H,K$) 
photometry of the Galactic Globular cluster \omcp. On the basis of the new data 
sets we addressed several questions concerning the properties of the sub-populations 
present in this stellar system. The main findings are the following:

{\em a)} {\em New calibrations of the $m_0, (\bmy)_0$ MIC relation.}
We have provided new empirical and semi-empirical calibrations of
$m_0, (\bmy)_0$ MIC relations. Photometric metallicities based on
the new MIC relations agree quite well with spectroscopic data
and with photometric metallicities based on similar MIC relations
available in the literature.

{\em b)} {\em Comparison between photometric and spectroscopic iron abundances.}
We performed a detailed comparison between photometric and
spectroscopic iron abundances. We selected iron abundances based
on high-resolution spectra collected by ND95 (39), P02 (8) and
J08 (118). We found,
using four different MIC relations, that the mean difference
between photometric and spectroscopic (ND95, P02) abundances is
$<\Delta\feh> \equiv  \feh_{phot}-\feh_{spec} = 0.17\pm0.01$
with a dispersion in the residuals of $\sigma$=0.31 dex (47 stars).
If we discard the RG stars affected by abundance
anomalies in  {\rm CN} or  {\rm CH}, the mean difference becomes
$<\Delta\feh>=$$0.02\pm0.02$ and the dispersion decreases by
a factor of two ($\sigma$=0.16 dex, 36 stars).
The same conclusion applies if we use the spectroscopic sample
by J08: the difference is $<\Delta\feh>=0.05\pm0.03$, with
$\sigma=0.37\,$dex (118 stars).  When we remove the RGs affected by
abundance anomalies we find that the mean difference deecreases 
($<\Delta\feh>=0.02\pm0.04$), while the dispersion in the residuals 
attains a similar value ($\sigma=0.37\,$dex, 93 stars). Note that 
measurements of the $S3839$ index are available for only a small 
fraction of the sample, but these suggest that photometric and 
spectroscopic abundances are, on average, minimally different. 
The dispersion of the residuals decreases once we remove all 
the RGs affected by abundance anomalies. These findings do not 
depend significantly on the adopted MIC relations, although the 
$m_0, (\bmy)_0$ MIC relations do have, on average, larger 
dispersions.

{\em c)} {\em Metallicity distributions based on different MIC relations.}
We have estimated the metallicity distribution using both empirical and
semi-empirical MIC relations. 
The six distributions based on two independent calibrations 
have similar properties, in particular, they show four main peaks at
$\feh_{phot}=$ $-1.73\pm0.08$, $-1.29\pm0.03$, $-1.05\pm0.02$,
$-0.80\pm0.04$, and three minor peaks at $-0.42\pm0.12$, $-0.07\pm0.08$ 
and $0.24\pm0.13$ dex (where the uncertainties are standard errors in 
the centroid of each peak). The four main
peaks agree quite well with low- (N96, SK96), medium-(H04, S05b) 
and high-resolution (J08) spectroscopic iron abundances and with previous 
photometric metallicities (HR00, S05a). Spectroscopic abundances also suggest the
occurrence of metal-rich ($\feh =-0.60\pm0.15$) stars in \omcp (ND95;
P00; Pancino 2004).  The stars belonging to the solar metallicity tail
should be regarded with caution since they might be either  {\rm CN}-strong stars
or field RGs.

{\em d)} {\em Identification of sub--populations according 
to their metal content.}
We identified four different metallicity regimes. The  weighted mean
fraction of stars in the {\em metal-poor}  ($\feh \le -1.49$) component
is $\sim39\pm1$\%, with 
a sharp cutoff for $\feh \approx -2.0$ dex.  In comparison, the
weighted mean fraction of stars in the {\em metal--intermediate}
($-1.49 < \feh \le -0.93$) component is  $\sim 32\pm 1$\%, while
the  fraction of {\em metal-rich}  ($-0.95 < \feh \le -0.15$) stars is
$\sim 19\pm4$\%,  and the solar metallicity stars ($\feh \approx 0$)
represent only a small fraction of the total ($\sim 8\pm5$\%). This last
apparent sub-population lacks a firm spectroscopic confirmation.

The new mosaic cameras available at telescopes of the 4--8m class,
and the increased sensitivity of new CCDs to short wavelengths 
combine to make medium-band photometry very promising for constraining 
the nature of any composite stellar population that may be found
in globular clusters. The same conclusion applies equally well to more 
complex systems such as the Galactic bulge and nearby dwarf galaxies 
(Faria et al.\ 2007).

{\em e)} {\em Abundance anomalies.}
We have cross-correlated our optical-NIR photometric catalog with
the large spectroscopic catalog collected by VL07. Using the
$S3839$ index (cyanogen band) defined by VL07 and 
$(y-K)_0$ we selected a sample of 181 candidate  {\rm CN}-strong
stars with a  {\rm CN} excess of $\delta{\hbox{CN}} > 0.2$. Taken at face 
value this selection would imply a fraction of  {\rm CN}-strong stars 
of the order of 50\%. We provided an independent estimate of the 
fraction of  {\rm CN}-strong stars from the difference between the 
iron content based on the semi-empirical $m_0, (\vmy)_0$ 
(more sensitive to the  {\rm CN}-band strength) and the semi-empirical 
$m_0, (\umy)_0$ relations. We found that the fraction of candidate 
 {\rm CN}-strong RGs ($\Delta\feh > 0.2$ dex) is of the order of 
$19\pm 1$\%.   
This fraction agrees quite well with the fraction of SGB 
stars that, according to Stanford et al.\ (2007), display 
enhancements in either {\rm C} ($\sim$ 17\%) or {\rm N} ($\sim$ 16\%) 
(see their Table~4). If these enhancements are caused by 
binary interaction, as suggested by Stanford et al.\ (2007, 
and references therein), the current evidence would imply that 
\omc hosts a population of binary stars that is at least a factor 
of five larger than suggested by spectroscopy (Mayor et al.\ 1996)
and a factor of two larger than the typical binary fraction  
believed to be present in most GCs (Davies et al.\ 2006). 

We also found that photometric metallicities correlate with 
the {\rm [Ca+Si/H]} abundance from $\feh \sim$ --1.6 up to Solar 
metallicity. More metal-poor stars display a plateau
at  {\rm [Ca+Si/H]}$\approx$--1.5 and  $\feh \lesssim$ --1.7 dex.   

This investigation is just a step in the ongoing effort to 
improve the use of Stroemgren indices as measures of chemical 
content. This should certainly be refined as more discriminating 
data become available. 
Our results also suggest the potential power of simultaneous 
use of optical data collected with other medium-band photometric 
systems. The DDO system (McClure \& van den Bergh 1968) with 
filters centered on the  {\rm CN} 
molecular band at $\lambda\sim 4215$ \AA, on the  G band near 
4300\,\AA, and on the magnesium complex near 5160\,\AA ~appears 
particularly promising. The use of the $hk$ index (ATT98) to 
estimate the calcium abundance of cluster HB and RG stars 
should also be explored further.
   

\acknowledgments
It is a real pleasure to thank J. van Loon for sending us his spectroscopic 
data in electronic form. 
We also thank S. Moehler for a detailed reading of an early draft of this 
manuscript and for several suggestions. 
We acknowledge an anonymous referee for his/her pertinent comments that 
helped us to improve the content and the readability of the manuscript.
We also acknowledge the second referee, Prof. B. Twarog, for his sharp 
and enlightening suggestions. 
This project was partially supported by the  grant Monte dei Paschi di 
Siena (P.I.: S. Degl'Innocenti), PRIN-INAF2006 (P.I.: M. Bellazzini), 
ASI-Project2006 (P.I.: F.R. Ferraro), PRIN-MIUR2007 (P.I.: G. Piotto). 
This publication makes use of data products from VizieR (Ochsenbein, Bauer,
\& Marcout 2000) and from the Two Micron All Sky Survey, which is a joint 
project of the University of Massachusetts and the Infrared Processing and 
Analysis Center/California Institute of Technology, funded by the National 
Aeronautics and Space Administration and the National Science Foundation.
We also thank the ESO and the HST Science Archive for their prompt support.



\begin{deluxetable}{lrc}
\tablewidth{0pt}
\tabletypesize{\scriptsize}
\tablecaption{Positional, photometric and structural parameters of the 
Galactic Globular Cluster \omcp.}\label{tbl-1}
\tablehead{
\colhead{Parameter}&
\colhead{         }&   
\colhead{Ref.\tablenotemark{a}} }
\startdata
$\alpha$ (J2000)                          &   13~26~45      &     1    \\  
$\delta$ (J2000)                          &  -47~28~37      &     1    \\  
$M_V$ (mag)\tablenotemark{b}              &    -10.3        &     1    \\  
$r_c$ (arcmin)\tablenotemark{c}           &  $2.58$      &     2    \\  
$r_t$ (arcmin)\tablenotemark{d}           &  $44.8$      &     2    \\  
$e$\tablenotemark{e}                      &  $0.15$         &     3    \\  
$\sigma_V$ (km~s$^{-1}$)\tablenotemark{f} &  $15$           &     4    \\  
E(B-V)\tablenotemark{g}                   &  $0.11\pm0.02$  &     5    \\  
$(m-M)_0$ (mag)\tablenotemark{h}          &  $13.70\pm0.12$ &     6    \\  
$\mu_{\alpha}\times cos\delta, \mu_{\delta}$ (mas/yr)\tablenotemark{i} & (-3.97, -4.38) &     7    \\  
Position angle (Degree)                   &  $100$          &     3    \\  
\enddata 
\tablenotetext{a}{References: 1) Harris 1996;
2) Trager, King \& Djorgovski 1995; 3) van de Ven et al.\ (2006);
4) Reijns et al.\ (2006); 5) Calamida et al.\ (2005);
6) Del Principe et al.\ (2006), Bono et al.\ (2008); 
7) van Leeuwen et al.\ (2000).} 
\tablenotetext{b}{Total visual magnitude.}   
\tablenotetext{c}{Core radius.}  
\tablenotetext{d}{Tidal radius.}   
\tablenotetext{e}{Eccentricity.}  
\tablenotetext{f}{Stellar central velocity dispersion.}   
\tablenotetext{g}{Mean cluster reddening.}   
\tablenotetext{h}{True distance modulus.} 
\tablenotetext{i}{Mean proper motion.} 
\end{deluxetable}

\begin{deluxetable}{cccrr}
\tablewidth{0pt}
\tabletypesize{\scriptsize}
\tablecaption{\strom data adopted in this investigation.}\label{tbl-2}
\tablehead{
\colhead{Field\tablenotemark{a}}&
\colhead{Images}&  
\colhead{FoV\tablenotemark{b}}& 
\colhead{RA\tablenotemark{c}}& 
\colhead{DEC\tablenotemark{c}}}
\startdata
LF & 25$u$,28$v$,22$b$,30$y$ & $13.7\times13.7$ & 13~26~24 & -47~28~12 \\
FG-1 & 3$u,3v,3b,3y$ & $13.7\times13.7$ & 13~26~24 & -47~28~12 \\
FG-2 & 4$u,4v,4b,4y$ & $13.7\times13.7$ & 13~25~48 & -47~38~24 \\
MH-1 & 2$v,2b,2y$ & $6.3\times6.3$ & 13~26~57 & -47~33~09 \\  
MH-2 & 3$v,3b,3y$ & $6.3\times6.3$ & 13~26~59 & -47~38~50 \\  
\enddata 
\tablenotetext{~}{NOTE. -- Table~2 is presented in its entirety in the 
electronic edition of the manuscript. A portion is shown here for 
guidance regarding its form and content.} 
\tablenotetext{a}{Data set collected with the 1.54m DANISH Telescope 
available at ESO (La Silla).}  
\tablenotetext{b}{Field of view (arcmin).}  
\tablenotetext{c}{Field coordinates (J2000): units of right ascension 
are hours, minutes, and seconds, while units of declination are degrees, 
arcminutes, and arcseconds.}
\end{deluxetable}

\begin{deluxetable}{lccrcc}
\tablewidth{0pt}
\tabletypesize{\scriptsize}
\tablecaption{Log of scientific CCD images of \omc collected by LF 
and by FG.}\label{tbl-3}
\tablehead{
\colhead{Frame}&
\colhead{Date\tablenotemark{a}}&
\colhead{HJD\tablenotemark{b}}&
\colhead{ET\tablenotemark{c}}&
\colhead{Filter\tablenotemark{d}}&
\colhead{Seeing\tablenotemark{e}} \\
\colhead{(1)}&
\colhead{(2)}&
\colhead{(3)}&
\colhead{(4)}&
\colhead{(5)}&
\colhead{(6)} } 
\startdata
           \multicolumn{6}{c}{LF run}\\  
dfsc0656 & 28-03-99 & 2451265.7938 & 450  & y &  1.32 \\ 
dfsc0660 & 28-03-99 & 2451265.8398 & 2000 & u &  2.59 \\ 
dfsc0661 & 28-03-99 & 2451265.8519 & 900  & v &  1.50  \\
dfsc0662 & 28-03-99 & 2451265.8600 & 600  & b &  1.58  \\
dfsc0663 & 28-03-99 & 2451265.8665 & 450  & y &  1.82  \\
\enddata 						     
\tablenotetext{~}{NOTE. -- Table~3 is presented in its entirety 
in the electronic edition of the manuscript. A portion is shown 
here for guidance regarding its form and content.}		     
\tablenotetext{a}{UT Date.}
\tablenotetext{b}{Heliocentric Julian Date.}
\tablenotetext{c}{Individual exposure time (sec).}
\tablenotetext{d}{The $u,v,b,y$ \strom filters of the 
1.5m Danish Telescope (ESO, La Silla).}
\tablenotetext{e}{Individual seeing (arcsec).} 
\end{deluxetable}

\begin{deluxetable}{lcccrcc}
\tablewidth{0pt}
\tabletypesize{\scriptsize}
\tablecaption{Log of scientific CCD images of \omc collected by MH.}\label{tbl-4}
\tablehead{
\colhead{Frame}&
\colhead{Field}&
\colhead{Date\tablenotemark{a}}&
\colhead{HJD\tablenotemark{b}}&
\colhead{ET\tablenotemark{c}}&
\colhead{Filter\tablenotemark{d}}&
\colhead{Seeing\tablenotemark{e}} \\
\colhead{(1)}&
\colhead{(2)}&
\colhead{(3)}&
\colhead{(4)}&
\colhead{(5)}&
\colhead{(6)}&
\colhead{(7)} } 
\startdata
           \multicolumn{7}{c}{MH 1993 run}    \\  
om1y3 & MH-1 & 13-05-93 & 2449120.6384 & 30 & y & 2.36\\
om1b3 & MH-1 & 13-05-93 & 2449120.6427 & 60 & b & 2.57\\
om1v3 & MH-1 & 13-05-93 & 2449120.6474 & 120 & v & 2.65\\
om2y3 & MH-2 & 13-05-93 & 2449120.7222 & 30 & y & 2.05\\
om2b3 & MH-2 & 13-05-93 & 2449120.7265 & 60 & b & 2.21\\
\enddata
\tablenotetext{~}{NOTE. -- Table~4 is presented in its entirety 
in the electronic edition of the manuscript. A portion is shown 
here for guidance regarding its form and content.}
\tablenotetext{a}{UT Date.}
\tablenotetext{b}{Heliocentric Julian Date.}   
\tablenotetext{c}{Individual exposure time (sec).}
\tablenotetext{d}{The $u,v,b,y$ \strom filters of the 1.54m Danish 
Telescope (ESO, La Silla).}
\tablenotetext{e}{Individual seeing (arcsec).} 
\end{deluxetable}

\begin{deluxetable}{lccrr} 
\tablewidth{0pt}
\tabletypesize{\scriptsize}
\tablecaption{NIR photometric data adopted in this investigation.}\label{tbl-5}
\tablehead{
\colhead{Field\tablenotemark{a}}&
\colhead{Images}&  
\colhead{FoV\tablenotemark{b}}& 
\colhead{RA\tablenotemark{c}}& 
\colhead{DEC\tablenotemark{c}} 
}
\startdata
ISAAC\_NE & 8$NB\_1.21$, 24$NB\_2.19$ & $2.5\times2.5$ & 13~26~36 & -47~27~31 \\
SOFI\_A & 36$J$,55$K_s$  & $4.9\times4.9$ & 13~26~46 & -47~28~36 \\
SOFI\_C & 33$u$,55$K_s$ & $4.9\times4.9$ & 13~26~07 & -47~24~25 \\
SOFI\_D & 12$J$,20$K_s$ & $4.9\times 4.9$ & 13~26~11 &  -47~34~52\\
SOFI\_om11 & 1$J,1K_s$ & $4.9\times 4.9$ & 13~27~08 &  -47~32~36\\  
 \enddata 
\tablenotetext{~}{NOTE. -- Table~5 is presented in its entirety in 
the electronic edition of the manuscript. A portion is shown here 
for guidance regarding its form and content.}			     
\tablenotetext{a}{Data set collected with the VLT Telescope available 
at ESO (Paranal) and with the NTT Telescope available at ESO (La Silla).}
\tablenotetext{b}{Field of view (arcmin)} 
\tablenotetext{c}{Field coordinates (J2000): units of right ascension are
hours, minutes, and seconds, while units of declination are degrees, 
arcminutes, and arcseconds.}
\end{deluxetable}

\clearpage 
\begin{deluxetable}{lccrcc}
\tablewidth{0pt}
\tabletypesize{\scriptsize}
\tablecaption{Log of scientific NIR CCD images of \omcp.}\label{tbl-6}
\tablehead{
\colhead{Frame}&
\colhead{Date\tablenotemark{a}}&
\colhead{HJD\tablenotemark{b}}&
\colhead{ET\tablenotemark{c}}&
\colhead{Filter\tablenotemark{d}}&
\colhead{Seeing\tablenotemark{e}} \\
\colhead{(1)}&
\colhead{(2)}&
\colhead{(3)}&
\colhead{(4)}&
\colhead{(5)}&
\colhead{(6)} } 
\startdata
                   \multicolumn{6}{c}{ISAAC\_NE field} \\
f1\_219\_01 & 2005-02-20  &  2453456.2889 & 66  & NB\_2.19  &  0.51\\
f1\_219\_02 & 2005-02-20  &  2453456.2851 & 66  & NB\_2.19  &  0.41\\
f1\_219\_03 & 2005-02-20  &  2453456.2839 & 66  & NB\_2.19  &  0.40\\
f1\_219\_04 & 2005-02-20  &  2453456.2801 & 66  & NB\_2.19  &  0.48\\
f1\_219\_05 & 2005-02-20  &  2453456.2789 & 66  & NB\_2.19  &  0.45\\
\enddata 						     
\tablenotetext{~}{NOTE. -- Table~6 is presented in its entirety in the 
electronic edition of the manuscript. A portion is shown here for 
guidance regarding its form and content.}			     
\tablenotetext{a}{UT Date.}				    
\tablenotetext{b}{Heliocentric Julian Date.} 
\tablenotetext{c}{Total exposure time --$ET=DIT*NDIT$-- (sec).}
\tablenotetext{d}{The $J,H,K_s$ NIR filters of the NTT Telescope 
(ESO, La Silla).} 
\tablenotetext{e}{Individual seeing (arcsec).} 
\end{deluxetable}

\begin{deluxetable}{llrc}
\tablewidth{0pt}
\tabletypesize{\scriptsize}
\tablecaption{Set of \strom standard stars observed by FG.}\label{tbl-7}
\tablehead{
\colhead{Star\tablenotemark{b}}&
\colhead{Date\tablenotemark{c}}&
\colhead{Filter}\\   
\colhead{(1)}&
\colhead{(2)}&
\colhead{(3)}
}
\startdata
HD112039 & 15-04-99 & $u,v,b,y$ \\ 
HD181720 & 15-04-99 & $u,v,b,y$ \\
HD167756 & 15-04-99 & $u,v,b,y$ \\
HD165896 & 15-04-99 & $u,v,b,y$ \\
HD165793 & 15-04-99 & $u,v,b,y$ \\
\enddata

\tablenotetext{~}{NOTE. -- Table~7 is presented in its entirety in 
the electronic edition of the manuscript. A portion is shown here 
for guidance regarding its form and content.}
\tablenotetext{a}{Standard stars have been selected from the catalogs 
of photometric standards of O93 and SN88. See text for more details.}
\tablenotetext{b}{Data set collected between April 1999 and June 1999 
with the 1.54m Danish Telescope available at ESO (La Silla).}   
\end{deluxetable}

\begin{deluxetable}{lcrr}
\tablewidth{0pt}
\tabletypesize{\scriptsize}
\tablecaption{Log of standard stars of the reference photometric night.}
\label{tbl-8}
\tablehead{
\colhead{Star\tablenotemark{a}}&
\colhead{HJD\tablenotemark{b}}&
\colhead{ET\tablenotemark{c}}&
\colhead{Filter}\\
\colhead{(1)}&
\colhead{(2)}&
\colhead{(3)}&
\colhead{(4)} } 
\startdata
HD119896 & 2451334.9467 &   7 &  y    \\    
HD119896 & 2451334.9471 &  15 &  b    \\   
HD119896 & 2451334.9475 &  25 &  v    \\   
HD119896 & 2451334.9480 &  70 &  u    \\   
HD119896 & 2451334.9895 &   7 &  y    \\   
\enddata
\tablenotetext{~}{NOTE. -- Table~8 is presented in its entirety in 
the electronic edition of the manuscript. A portion is shown here 
for guidance regarding its form and content.}  
\tablenotetext{a}{HD standard stars have been observed during the 
night of June the 6th, 1999, with the 1.54m Danish Telescope 
available at ESO (La Silla).}
\tablenotetext{b}{Heliocentric Julian Date.}
\tablenotetext{c}{Individual exposure time (sec).}		  
\end{deluxetable}

\begin{deluxetable}{lccrc}
\tablewidth{0pt}
\tabletypesize{\scriptsize}
\tablecaption{Log of HD standard stars observed at different airmass values.\tablenotemark{a}}
\label{tbl-9}
\tablehead{
\colhead{Star}&
\colhead{Airmass\tablenotemark{b}}&
\colhead{HJD\tablenotemark{c}}&
\colhead{ET\tablenotemark{d}}&
\colhead{Filter}\\
\colhead{(1)}&
\colhead{(2)}&
\colhead{(3)}&
\colhead{(4)}&
\colhead{(5)}}
\startdata
HD119896  & 1.263 & 51334.9467 & 7  &  y \\	 
HD119896  & 1.262 & 51334.9471 & 15 &  b   \\	
HD119896  & 1.260 & 51334.9475 & 25 &  v   \\	
HD119896  & 1.258 & 51334.9480 & 70 &  u   \\	
HD119896  & 1.134 & 51334.9895 & 7  &  y \\	
HD119896  & 1.133 & 51334.9898 & 15 &  b   \\	
HD119896  & 1.132 & 51334.9902 & 25 &  v   \\	
HD119896  & 1.132 & 51334.9908 & 70 &  u   \\	
HD119896  & 1.068 & 51335.0353 & 7  &  y \\	
HD119896  & 1.068 & 51335.0357 & 15 &  b   \\	
HD119896  & 1.067 & 51335.0361 & 25 &  v   \\  
HD119896  & 1.065 & 51335.0366 & 70 &  u   \\	
\enddata
\tablenotetext{a}{HD standard stars observed during the reference
night, June the 6th, 1999, with the 1.54m Danish Telescope available 
at ESO (La Silla).}   
\tablenotetext{b}{Individual airmass value.}  
\tablenotetext{c}{Heliocentric Julian Date.}   
\tablenotetext{d}{Individual exposure time (sec).}  		   
\end{deluxetable}

\begin{deluxetable}{lcccc}
\tablewidth{0pt}
\tabletypesize{\scriptsize}
\tablecaption{Extinction coefficients estimated for the observing
 nights.\tablenotemark{a}}\label{tbl-10}
\tablehead{
\colhead{Date}&
\colhead{$k_u$}&
\colhead{$k_v$}&
\colhead{$k_b$}&
\colhead{$k_y$}\\
\colhead{(1)}&
\colhead{(2)}&
\colhead{(3)}&
\colhead{(4)}&
\colhead{(5)}}
\startdata
15-04-99 & 0.230 & 0.130 & 0.074 & 0.064 \\	 
16-04-99 & 0.350 & 0.308 & 0.187 & 0.127 \\	 
19-04-99 & 0.522 & 0.309 & 0.185 & 0.145 \\	 
22-04-99 & 0.412 & 0.273 & 0.162 & 0.120 \\	 
06-06-99 & 0.604 & 0.403 & 0.270 & 0.176 \\  
\enddata
\tablenotetext{a}{Extinction coefficients estimated using  
a set of HD standard stars observed at different airmass 
values.}  
\end{deluxetable}				   

\begin{deluxetable}{lccccccc}
\tablewidth{0pt}
\tabletypesize{\scriptsize}
\tablecaption{Multilinear regression coefficients for the \strom  
MIC relations: $m = \alpha + \beta\cdot\feh + \gamma\cdot CI +
\delta\cdot(\feh\cdot CI) + \epsilon\cdot CI^2 + \zeta\cdot(\feh\cdot CI^2)$.\label{tbl-11}}
\tablehead{
\colhead{Relation}&
\colhead{$\alpha$}&
\colhead{$\beta$}&
\colhead{$\gamma$}&
\colhead{$\delta$}&
\colhead{$\epsilon$}&
\colhead{$\zeta$}&
\colhead{$r$}\\
\colhead{(1)}&
\colhead{(2)}&
\colhead{(3)}&
\colhead{(4)}&
\colhead{(5)}&
\colhead{(6)}&
\colhead{(7)}&
\colhead{(8)}}
\startdata
\multicolumn{8}{c}{Empirical based on selected GCs (see CA07)}  \\  
$m_0, (\bmy)_0$ & -0.51 & -0.17$\pm$0.04 & 1.66$\pm$0.20 & 0.56$\pm$0.13 & -0.020$\pm$0.14 & -0.003$\pm$0.090 & 1.00 \\    
$[m], (\bmy)_0$ & -0.47 & -0.15$\pm$0.05 & 1.90$\pm$0.21 & 0.53$\pm$0.13 &  0.001$\pm$0.14 &  0.004$\pm$0.090 & 1.00 \\    
\multicolumn{8}{c}{Semi-empirical based on transformations by Clem et al.\ (2004)} \\  
$m_0, (\bmy)_0$ & -0.64 & -0.22$\pm$0.06 & 2.04$\pm$0.30 & 0.71$\pm$0.18 & -0.22$\pm$0.20 & -0.10$\pm$0.12 & 0.99 \\    
$[m], (\bmy)_0$ & -0.51 & -0.16$\pm$0.06 & 2.01$\pm$0.26 & 0.57$\pm$0.16 & 0.003$\pm$0.18 & 0.003$\pm$0.110 & 0.99 \\    
\enddata
\tablenotetext{a}{Multi correlation coefficient.}   

\end{deluxetable}				   


\begin{deluxetable}{lccccccccc}
\tablewidth{0pt}
\tablecaption{Spectroscopic measurements from Rutledge et al.\ (1997)
and photometric metallicity estimates for the GCs adopted to validate
the MIC relations.\label{tbl-12}}
\tablehead{
\colhead{Relation}&
\colhead{M92}&
\colhead{NGC6397}&
\colhead{M13}&
\colhead{NGC6752}&
\colhead{NGC288}&
\colhead{NGC1851}&
\colhead{NGC362}&
\colhead{M71}&
\colhead{NGC104}\\
\colhead{(1)}&
\colhead{(2)}&
\colhead{(3)}&
\colhead{(4)}&
\colhead{(5)}&
\colhead{(6)}&
\colhead{(7)}&
\colhead{(8)}&
\colhead{(9)}&
\colhead{(10)}}
\startdata
\multicolumn{10}{c}{Spectroscopy\tablenotemark{a}}  \\
$\ldots$ & -2.24\tablenotemark{b} $\pm$0.10 & -1.91$\pm$0.14 & -1.65$\pm$0.06 & -1.54$\pm$0.09 & -1.40$\pm$0.12 & -1.33$\pm$0.10 & -1.27$\pm$0.07 & -0.73\tablenotemark{c} $\pm$0.05 & -0.71$\pm$0.05 \\
\multicolumn{10}{c}{Empirical based on selected GCs}  \\
$m_0,(\bmy)_0$ & $\ldots$ & -2.04$\pm$0.15 & $\ldots$ & -1.67$\pm$0.18 & -1.30$\pm$0.11 & $\ldots$ & -1.24$\pm$0.30 & -0.48$\pm$0.34 & $\ldots$ \\
$[m],(\bmy)_0$ & $\ldots$ & -2.02$\pm$0.15 & $\ldots$ & -1.67$\pm$0.16 & -1.31$\pm$0.12 & $\ldots$ & -1.28$\pm$0.24 & -0.55$\pm$0.32 & $\ldots$ \\
\multicolumn{10}{c}{Semi-empirical based on transformations by Clem et al.\ (2004)} \\
$m_0,(\bmy)_0$ & -2.11$\pm$0.17 &-2.05$\pm$0.16 & -1.79$\pm$0.11 & -1.69$\pm$0.15 & -1.32$\pm$0.11 & -1.53$\pm$0.18 & -1.27$\pm$0.25 & -0.50$\pm$0.36 & -0.48$\pm$0.38 \\
$[m],(\bmy)_0$ & -2.09$\pm$0.16 &-2.05$\pm$0.15 & -1.81$\pm$0.09 & -1.72$\pm$0.15 & -1.38$\pm$0.11 & -1.55$\pm$0.18 & -1.34$\pm$0.22 & -0.65$\pm$0.31 & -0.61$\pm$0.31 \\
\enddata
\tablenotetext{a} {Cluster metal abundances according to Rutledge et al.\ (1997)
in the metallicity scale by Zinn \& West (1984) and Zinn (1985).}
\tablenotetext{b} {Value from Zinn \& West (1984).}
\tablenotetext{c} {Value from the GC catalog by Harris (1996).}

\end{deluxetable} 


\begin{deluxetable}{clcccccc}
\tablewidth{0pt}
\tabletypesize{\scriptsize}
\tablecaption{Peaks and sigmas of multigaussian fits to the RG metallicity 
distributions based on the different MIC relations.}\label{tbl-13}
\tablehead{
\colhead{Relation}&
\colhead{MP\tablenotemark{a}}&
\colhead{MI1\tablenotemark{b}}&
\colhead{MI2\tablenotemark{b}}&
\colhead{MR1\tablenotemark{c}}&
\colhead{MR2\tablenotemark{c}}&
\colhead{SM1\tablenotemark{d}}&
\colhead{SM2\tablenotemark{d}}
}
\startdata
\multicolumn{8}{c}{Empirical calibrations}  \\
$m_0,(\umy)_0$ & -1.74/0.32 & -1.26/0.09 & -1.03/0.11 & -0.81/0.11 & -0.48/0.24 & -0.07/0.21 & 0.32/0.09 \\
$m_0,(\vmy)_0$ & -1.70/0.32 & -1.29/0.10 & -1.06/0.07 & -0.79/0.29 & -0.20/0.11 & -0.02/0.02 & 0.16/0.22 \\
$m_0,(\bmy)_0$ & -1.79/0.26 & -1.30/0.15 & -1.04/0.09 & -0.79/0.16 & -0.46/0.15 & -0.06/0.21 & 0.35/0.10 \\
$Average$\tablenotemark{e}      & -1.72/0.31 & -1.30/0.09 & -1.09/0.09 & -0.88/0.10 & -0.61/0.22 & -0.03/0.17 & 0.32/0.08 \\
\multicolumn{8}{c}{Semi-empirical calibrations} \\
$m_0,(\umy)_0$ & -1.70/0.25 & -1.32/0.09 & -1.10/0.09 & -0.89/0.11 & -0.55/0.15 & -0.20/0.07 & 0.00/0.15 \\
$m_0,(\vmy)_0$ & -1.59/0.28 & -1.24/0.10 & -1.03/0.10 & -0.76/0.16 & -0.33/0.22 & 0.05/0.08 & 0.33/0.13 \\
$m_0,(\bmy)_0$ & -1.84/0.24 & -1.33/0.20 & -1.03/0.18 & -0.79/0.19 & -0.49/0.14 & -0.10/0.17 & 0.31/0.11 \\
$Average$\tablenotemark{e}      & -1.69/0.26 & -1.28/0.12 & -1.07/0.10 & -0.88/0.04 & -0.71/0.26 & -0.03/0.18 & 0.30/0.17 \\
               &            &            &            &            &            &            &           \\
$Mean$\tablenotemark{f}& -1.73$\pm$0.08  & -1.29$\pm$0.03 & -1.05$\pm$0.03 & -0.80$\pm$0.04 & -0.42$\pm$0.13 & -0.07$\pm$0.08 & 0.24$\pm$0.13 \\ 
\enddata
\tablenotetext{a}{Peak and sigma in the metal-poor regime (MP, \feh $\le -1.49$ dex).}
\tablenotetext{b}{Peaks and sigmas in the metal-intermediate regime (MI, -1.49$<$\feh $\le$-0.93 dex).}
\tablenotetext{c}{Peaks and sigmas in the metal-rich regime (MR, -0.93$<$\feh$\le$-0.19 dex).}
\tablenotetext{d}{Peaks and sigmas in the solar metallicity regime (SM, \feh$>$-0.19 dex).}
\tablenotetext{e}{Average either of empirical or semi-empirical calibrations (see Fig.~17).}  
\tablenotetext{f}{Weighted mean peak values of the six MIC relations. The uncertainties are the 
errors on the mean. The typical uncertainty in the position of the individual peaks is $\pm$ 0.04 dex, 
i.e. the bin size adopted in the metallicity distributions.}
\end{deluxetable}


\begin{deluxetable}{clccr}
\tablewidth{0pt}
\tablecaption{Relative fraction of \omc RGs in the selected metallicity groups.}\label{tbl-14}
\tablehead{
\colhead{Relation}&
\colhead{MP\tablenotemark{a}}&
\colhead{MI\tablenotemark{b}}&
\colhead{MR\tablenotemark{c}}&
\colhead{SM\tablenotemark{d}}
}
\startdata
\multicolumn{5}{c}{Empirical calibrations}\\
$m_0,(\umy)_0$ & 47$\pm$2\tablenotemark{e} & 31$\pm$1 & 16$\pm$1 & 5$\pm$1 \\
$m_0,(\vmy)_0$ & 38$\pm$1 & 32$\pm$1 & 23$\pm$1 & 7$\pm$1 \\
$m_0,(\bmy)_0$ & 38$\pm$1 & 31$\pm$1 & 17$\pm$1 & 14$\pm$1 \\
$Average$\tablenotemark{f}      & 40$\pm$1 & 31$\pm$1 & 20$\pm$1 & 9$\pm$1 \\
\multicolumn{5}{c}{Semi-empirical calibrations}\\
$m_0,(\umy)_0$ & 48$\pm$2 & 33$\pm$1 & 16$\pm$1 & 3$\pm$1 \\
$m_0,(\vmy)_0$ & 36$\pm$1 & 34$\pm$1 & 25$\pm$1 & 7$\pm$1 \\
$m_0,(\bmy)_0$ & 36$\pm$1 & 33$\pm$1 & 17$\pm$1 & 14$\pm$1 \\
$Average$\tablenotemark{f}      & 38$\pm$1 & 34$\pm$1 & 20$\pm$1 & 8$\pm$1 \\
               &          &          &          &         \\
$Mean$\tablenotemark{g}& 39$\pm$1 & 32$\pm$1 & 19$\pm$4 & 8$\pm$5 \\
\enddata
\tablenotetext{a}{Fraction of metal-poor RGs (MP, \feh $\le -1.49$).}
\tablenotetext{b}{Fraction of metal-intermediate RGs (MI, -1.49$<$\feh $\le$-0.93).}
\tablenotetext{c}{Fraction of metal-rich RGs (MR, -0.93$<$\feh$\le$-0.19).}
\tablenotetext{d}{Fraction of solar metallicity RGs  (SM, \feh$>$-0.19).}
\tablenotetext{e}{One sigma Poisson error on the relative fractions.}
\tablenotetext{f}{Average either of empirical or of semi-empirical calibrations (see Fig.~17).}  
\tablenotetext{g}{Weighted mean values of the six MIC relations.}
\end{deluxetable}

\end{document}